\documentclass[%
 reprint,
 superscriptaddress,
 preprintnumbers,
nofootinbib,
 amsmath,amssymb,
 aps,
]{revtex4-2}

\usepackage{graphicx}
\usepackage{subcaption}
\usepackage{orcidlink}
\usepackage{dcolumn}
\usepackage{bm}
\usepackage{hyperref}
\hypersetup{
	colorlinks=true,
	linkcolor=blue,
	filecolor=magenta,      
	urlcolor=blue,
	citecolor=magenta
}
\usepackage{amsmath,amssymb,mathtools}
\usepackage{graphics,graphicx,tikz,tensor}
\usetikzlibrary{arrows,decorations.pathmorphing,decorations.markings,patterns,arrows.meta}
\usetikzlibrary{positioning, shapes, snakes, arrows}
\usepackage{soul}

\tikzset{
	graviton/.style={line width=.8pt, -latex,decorate, decoration={snake, segment length=4pt,amplitude=1.8pt, pre length=.1cm, post length=.25cm}},
	worldline/.style={gray, line width=1pt},
	worldlineBold/.style={black, line width=.6pt},
	zUndirected/.style={line width=1pt},
	zParticle/.style={line width=1pt,postaction={decorate},decoration={markings,mark=at position .6 with {\arrow[#1]{latex}}}},
	zParticleF/.style={line width=1pt,postaction={decorate}},
	cscalar/.style={line width=1pt,postaction={decorate},decoration={markings,mark=at position .6 with {\arrow[#1]{latex}}}},
	cscalar2/.style={line width=1pt,postaction={decorate},decoration={markings,mark=at position .8 with {\arrow[#1]{latex}}}},
	photon/.style={line width =.8pt, decorate, decoration={snake, segment length=4pt, amplitude=1.8pt,  pre length=.1cm, post length=.1cm}}
}

\def\Eqn#1{Eq.~\eqref{#1}}

\def\Rcite#1{Ref.~\cite{#1}}
\def\Rcites#1{Refs.~\cite{#1}}

\begin{document}

\preprint{HU-EP-26/30}
\preprint{QMUL-PH-26-33} 

\title{Angular Momentum Loss at Fourth Post-Minkowskian Order}

\author{Alessandro Georgoudis\,\orcidlink{0000-0002-1601-4218}}
\affiliation{School of Physical and Chemical Sciences,
  Queen Mary University of London, London E1~4NS, United Kingdom}
\affiliation{Department of Mathematics, City St George's, University of London, London EC1V~0HB, United Kingdom}
\author{Carlo Heissenberg\,\orcidlink{0000-0001-5775-9526}}
\affiliation{Institut de Physique Th\'eorique, CEA, CNRS, Universit\'e Paris-Saclay, F-91191, Gif-sur-Yvette Cedex, France}
	
\author{Gustav Mogull\,\orcidlink{0000-0003-3070-5717}}
\affiliation{School of Mathematical Sciences, Queen Mary University of London, London, E1 4NS, United Kingdom}
\affiliation{Institut f\"ur Physik, Humboldt Universit\"at zu Berlin, 10099 Berlin, Germany}
\affiliation{Max Planck Institute for Gravitational Physics (Albert Einstein Institute), Am M\"uhlenberg 1, 14476 Potsdam, Germany}

\author{Rodolfo Russo\,\orcidlink{0000-0003-4732-6369}}
\affiliation{School of Mathematical Sciences, Queen Mary University of London, London, E1 4NS, United Kingdom}

\begin{abstract}
We calculate the total loss of angular momentum for a gravitational scattering of two non-spinning point-like objects at $\mathcal{O}(G^4)$.
Our methodology combines use of both amplitudes and worldline-based methods.
We derive the integrand from two copies of the gravitational waveform
with cuts on massive lines arising naturally from the worldline.
We then glue these two copies and evaluate the resulting phase-space integrals using reverse unitarity,
in the amplitudes spirit. 
We obtain separately the radiated angular momentum and the static contribution that, at this  
order, is entirely determined by the initial shear and the linear memory. Combining these two terms, we derive relativistic expressions for the angular momentum loss 
and for the infrared-finite component of the mass dipole loss, which in the low-velocity
limit match results available in the literature.
We also use this result to derive the conservative  
deflection angle at $\mathcal{O}(G^5)$ and first self-force order.
\end{abstract}

\maketitle

Diagrammatic methods inspired by particle physics have triggered a revolution
in the gravitational two-body scattering problem by providing powerful tools for
calculating observable quantities.
The premise that the two bodies' size,
which for black holes is characterized by their Schwarzschild radii,
is much smaller than the impact parameter $b$
invites a pointlike approximation in which one expands in $Gm_i/b\ll1$,
$G$ being Newton's constant and $m_i$ being the masses.
This post-Minkowskian (PM) expansion in powers of $G$
makes no assumption on the velocities of
the two bodies, unlike the post-Newtonian (PN) that also assumes slow velocities.

An important breakthrough achieved using particle-based methods
was the calculation of the 3PM ($G^3$)
deflection angle obtained using scattering amplitudes~\cite{Bern:2019nnu,Bern:2019crd}.
In parallel with amplitudes-based methods,
worldline-based methods have been crucial for pushing to higher PM orders
due to their efficient targeting of the classical $\hbar\to0$ limit
\cite{Kalin:2020mvi,Kalin:2020fhe,Mogull:2020sak}.
The high-energy behavior of the 3PM result motivated further investigations, performed both with amplitude and worldline techniques, which led to a deeper understanding of dissipative effects due to radiation-reaction \cite{DiVecchia:2020ymx,Damour:2020tta,DiVecchia:2021ndb} and energy losses \cite{Herrmann:2021lqe,DiVecchia:2021bdo,Herrmann:2021tct,Bjerrum-Bohr:2021vuf,Bjerrum-Bohr:2021din,Brandhuber:2021eyq,Kalin:2022hph,Jakobsen:2022psy} at 3PM order.
The state of the art then rapidly advanced to 4PM for both deflection
and radiated energy-momentum \cite{Bern:2021dqo,Dlapa:2021npj,Bern:2021yeh,Dlapa:2021vgp,Dlapa:2022lmu,Jakobsen:2023ndj,Jakobsen:2023hig,Damgaard:2023vnx,Damgaard:2023ttc},
and more recent work reaches as far as 5PM order
\cite{Driesse:2024xad,Driesse:2024feo,Driesse:2026qiz,Bern:2024adl,Bern:2025wyd}.
Amplitudes and worldlines have advanced in parallel so far,
but in this work we combine the simplicity of the integrands
and integral families arising naturally from the worldline with
powerful amplitude tools for phase-space integration and soft theorems.

Two crucial observables involved in the two-body scattering
are the energy and angular momentum losses,
$E_{\rm loss}$ and $J_{\rm loss}$, due to interactions with the gravitational field,
$h_{\mu\nu} = g_{\mu\nu}-\eta_{\mu\nu}$.
The latter has the form
\begin{equation}
    \label{eq:Jtd}
    J_\text{loss} =
	-\frac{1}{32\pi G}  \lim_{r\to\infty} \int\!{\rm d}u \oint r^2d\Omega \, \partial_u h_{AB} \partial_\phi h^{AB}\,,
\end{equation}
where $u$ is the retarded time parametrizing,
together with the angles $A,B=\theta, \phi$, future null infinity.
Compared with $E_{\rm loss}$,
which can be easily inferred from the change in mechanical linear momentum,
$J_{\rm loss}$ has proved more difficult to calculate explicitly.
As evidence of this,
the previous 3PM state of the art for $J_{\rm loss}$
lags far behind the aforementioned 5PM for other observables.

Part of the challenge when computing $J_{\rm loss}$ is that
it contains both dynamical (radiative) and static contributions,
unlike $E_{\rm loss}$ that contains only dynamical terms.
In the intrinsic Bondi--Metzner--Sachs (BMS) frame \cite{Veneziano:2022zwh}, static contributions 
arise due to interference between the initial shear and the gravitational memory,
so $J_{\rm loss}$ begins at 2PM~\cite{Damour:2020tta}
while $E_{\rm loss}$ begins at 3PM.
However, in \Rcite{DiVecchia:2022owy} the eikonal approach 
was used to provide a non-perturbative formula for the static contributions due to the linear memory.
Ref.~\cite{Heissenberg:2024umh} also clarified that interference terms involving
non-linear memory do not contribute to $J_{\rm loss}$ below 5PM order~\cite{Georgoudis:2025vkk}.
Calculations of the remaining dynamical contributions involve the same family
of loop integrals involved in $E_{\rm loss}$,
enabling the 3PM calculations in Refs.~\cite{Manohar:2022dea,DiVecchia:2022piu,Mogull:2025cfn,Jakobsen:2026ayw}.
So far, at 4PM only the odd-in-velocity part of $J_{\rm loss}$ has been obtained,
by rewriting it in terms of a two-loop calculation~\cite{Heissenberg:2025ocy}.

In this Letter, we complete the calculation of $J_{\rm loss}$ at $\mathcal{O}(G^4)$
by deriving its most challenging contribution, which is even in velocity.
Like $E_{\rm loss}$, $J_{\rm loss}$ admits a simple functional
representation involving only logarithms and dilogarithms.
Furthermore, when combined with linear response \cite{Bini:2025zvk},
this result enables us to construct a conservative result for the 5PM scattering angle
up to linear order in self-force,
by subtracting radiative fluxes from the complete angle~\cite{Driesse:2024feo}.
Finally, we obtain the 4PM expression for the mass dipole loss along the impact parameter, which is free of infrared ambiguities.
 
\paragraph*{\textbf{Radiated Energy and Angular Momentum.}}
Here we calculate the radiative contribution $\boldsymbol{J}^{\alpha\beta}$
to the full angular momentum loss $J^{\alpha\beta}=\boldsymbol{J}^{\alpha\beta}+\mathcal{J}^{\alpha\beta}$, where $\mathcal{J}^{\alpha\beta}$ is the static contribution to be discussed below.
Our main focus is on the spacelike component 
\begin{align}\label{eq:jradjstat}
	\boldsymbol{J}_\text{rad} = b_\alpha \boldsymbol{J}^{\alpha\beta} p_\beta/(pb)\,,
\end{align} 
$b^\alpha$ being the impact parameter of the scattering process and
$p^\alpha$ being the spatial momentum of particle 1 in the center-of-mass frame~\eqref{eq:pCM}.
As a useful cross-check of our methodology,
we also compute the complete energy-momentum loss $P^\alpha$, which lacks a static contribution.

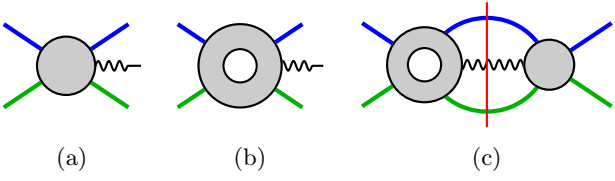
\begin{figure}[t!]
    \begin{subfigure}[b]{0.26\columnwidth}
		\centering
		\begin{tikzpicture}[scale=.55]
        \draw[thick,white] (0,1.5)--(0,-1.5);
    \draw[photon] (0,0)  -- (1.8,0);
			\draw [ultra thick, blue] (-1.5,1) -- (0,0);
			\draw [ultra thick, blue] (0,0) -- (1.5,1);
			\draw [ultra thick, green!70!black] (-1.5,-1) -- (0,0);
			\draw [ultra thick, green!70!black] (0,0) -- (1.5,-1);
			\filldraw[black!20!white, thick] (0,0) ellipse (.7 and .7);
			\draw[thick] (0,0) ellipse (.7 and .7);
		\end{tikzpicture}
		\caption{}
		\label{fig:tree23}
	\end{subfigure}
    \begin{subfigure}[b]{0.26\columnwidth}
		\centering
		\begin{tikzpicture}[scale=.55]
        \draw[thick,white] (0,1.5)--(0,-1.5);
    \draw[photon] (0,0)  -- (2,0);
			\draw [ultra thick, blue] (-1.5,1) -- (0,0);
			\draw [ultra thick, blue] (0,0) -- (1.5,1);
			\draw [ultra thick, green!70!black] (-1.5,-1) -- (0,0);
			\draw [ultra thick, green!70!black] (0,0) -- (1.5,-1);
			\filldraw[black!20!white, thick] (0,0) ellipse (1 and 1);
			\draw[thick] (0,0) ellipse (1 and 1);
			\filldraw[white, thick] (0,0) ellipse (.4 and .4);
			\draw[thick] (0,0) ellipse (.4 and .4);
		\end{tikzpicture}
		\caption{}
		\label{fig:oneloop23}
	\end{subfigure}
    \begin{subfigure}[b]{0.44\columnwidth}
    		\centering
			\begin{tikzpicture}[scale=.55]
				\draw [ultra thick, blue] (-3,1) -- (-1.5,0);
				\draw [ultra thick, blue] (-1.5,0) .. controls (-1,1.5) and (1,1.5) .. (1.5,0);
				\draw [ultra thick, blue] (3,1) -- (1.5,0);
				\draw[photon] (-1.5,0)  -- (2,0);
				\draw [ultra thick, green!70!black] (-3,-1) -- (-1.5,0);
				\draw [ultra thick, green!70!black] (-1.5,0) .. controls (-1,-1.5) and (1,-1.5) .. (1.5,0);
				\draw [ultra thick, green!70!black] (3,-1) -- (1.5,0);
				\filldraw[black!20!white, thick] (-1.5,0) ellipse (.9 and .9);
				\draw[thick] (-1.5,0) ellipse (.9 and .9);
				\filldraw[white, thick] (-1.5,0) ellipse (.4 and .4);
				\draw[thick] (-1.5,0) ellipse (.4 and .4);
				\filldraw[black!20!white, thick] (1.5,0) ellipse (.6 and .6);
				\draw[thick] (1.5,0) ellipse (.6 and .6);
                \draw[thick,red] (0,1.5)--(0,-1.5);
			\end{tikzpicture}
		\caption{}
		\label{fig:threePC3loop}
	\end{subfigure}
		\caption{\small Tree-level/one-loop ingredients, and their gluing.
		}
		\label{fig:tree1loopblobs}
\end{figure}

Both $P^\alpha$ and $\boldsymbol{J}^{\alpha\beta}$ arise due to dynamically propagating gravitational waves,
admitting a simple representation
in terms of the waveform $\tilde{W}_{\mu\nu}$~\cite{Herrmann:2021lqe,Manohar:2022dea,DiVecchia:2022owy}:
\begin{subequations}\label{eq:representationsW}
	\begin{align}
	P^\alpha &= \int_k k^\alpha \tilde{W}^\ast \tilde{W}\,,\label{eq:energymomentumWW}\\
	\boldsymbol{J}_{\alpha\beta} &=-i\int_k \bigg( \tilde{W}^\ast k^{\phantom{\dfrac{\mu}{\mu}}}_{[\alpha} \frac{\partial\tilde{W}}{\partial k^{\beta]}}
	+
	2 \tilde{W}^{\ast\phantom{\mu}}_{\mu [\alpha}\tilde W^{\mu}_{\beta]}
	\bigg)\,,\label{eq:angularmomentumWW}
\end{align}\end{subequations}
where $\int_k = \int 2\pi\theta(k^0)\delta(k^2) \frac{d^Dk}{(2\pi)^D}$ is the on-shell integration over the graviton's momentum in $D=4-2\epsilon$ dimensions.
Here and in the following, contraction with $\eta_{\mu\rho}\eta_{\nu\sigma}-\frac{1}{D-2}\eta_{\mu\nu}\eta_{\rho\sigma}$ is implicit. Both representations in \Eqn{eq:representationsW} rely on the Ward identity:
\begin{equation}\label{eq:WardIdentity}
	k_\mu \tilde{W}^{\mu\nu}= 0\,,
\end{equation}
which is a consequence of gauge-invariance.
We can write the waveform up to next-to-leading order \cite{Kovacs:1977uw,Kovacs:1978eu,Luna:2017dtq,Jakobsen:2021smu,Mougiakakos:2021ckm,DeAngelis:2026kew,Brandhuber:2023hhy,Herderschee:2023fxh,Elkhidir:2023dco,Georgoudis:2023lgf,Caron-Huot:2023vxl,Georgoudis:2023ozp,Georgoudis:2023eke,Georgoudis:2024pdz,Bini:2024rsy,Brunello:2025eso} as the Fourier transform \eqref{eq:inelasticFT},
denoted by a tilde, of a waveform kernel
\begin{equation}\label{eq:waveformkernel}
	W^{\mu\nu} = \mathcal{A}_0^{\mu\nu} + \left[\mathcal{B}_1^{\mu\nu} + i\,\mathcal{C}_1^{\mu\nu}\right]+\mathcal{O}(G^{7/2})\,,
\end{equation}
where $\mathcal{A}_0$ is the $\mathcal{O}(G^{3/2})$ classical $2\to3$ amplitude at tree level (Fig.~\ref{fig:tree23}), and the square brackets enclose the $\mathcal{O}(G^{5/2})$ one-loop kernel (Fig.~\ref{fig:oneloop23}), with $\mathcal{B}_1$ the real part and $\mathcal{C}_1$ the imaginary part. 

We consider a basis consisting of the particles initial velocities $u_a^\alpha$, with $a=1,2$, and the relative impact parameter $b^\alpha=b_1^\alpha-b_2^\alpha$,
where $\gamma=-u_1\cdot u_2$ and\footnote{The distinction between the \emph{initial} $u_a^\mu$, $b^\alpha$ and the \emph{eikonal} ones, $\tilde{u}_a^\alpha \simeq u_a^\alpha + \mathcal{O}(G) b^\alpha$ and $b_e^\alpha = b^\alpha + \mathcal{O}(G) u_{a}^\alpha$, which are symmetric in the conservative limit, is irrelevant for $u_a^\mu P_\mu$ and $u_a^\mu \boldsymbol{J}_{\mu\nu} b^\nu$ at $\mathcal{O}(G^4)$. This is because $b^\mu P_\mu$ and $u_1^\mu \boldsymbol{J}_{\mu\nu}u_2^\nu$ vanish at $\mathcal{O}(G^3)$. Consistently, this change of basis is equivalent to altering $\mathcal{C}_1^{\mu\nu}$ in \eqref{eq:waveformkernel}, to which $u_a^\mu P_\mu$ and $u_a^\mu \boldsymbol{J}_{\mu\nu} b^\nu$ are insensitive.} $u_a\cdot b=0$.
The radiated momentum is then decomposed as
\begin{equation}\label{eq:Pdecomp}
	P^\alpha = \sum_{a=1,2}(\check u_a^\mu P_\mu) u_a^\alpha + (b^\mu P_\mu) b^\alpha/b^2\,,
\end{equation}
where $\check u_a$ are the dual velocities satisfying $u_a\cdot\check u_b=\delta_{ab}$.
The radiated energy in the center-of-mass frame reads
\begin{equation}
	E_\text{loss} = - t_\alpha P^\alpha\,,\qquad
	t^\alpha = (m_1u_1^\alpha + m_2 u_2^\alpha)/E\,,
\end{equation}
where $E^2=m_1^2+2 m_1 m_2\gamma+m_2^2$ so that $t^2=-1$.
Meanwhile, the radiated angular momentum transforms under translations by $a^\alpha$ according to
\begin{equation}\label{eq:translationJ}
	\boldsymbol{J}^{\alpha\beta} \mapsto \boldsymbol{J}^{\alpha\beta} + a^{[\alpha} P^{\beta]}\,,
\end{equation}
while the static loss $\mathcal{J}^{\alpha\beta}$ is invariant. 
Fixing for now the translation frame by placing particle $2$ in the origin of the transverse plane, $b^\alpha_2=0$, we can then decompose  $\boldsymbol{J}^{\alpha\beta}$ as
\begin{equation}\label{eq:Jdecomp}
	\boldsymbol{J}^{\alpha\beta} =\!\! \sum_{a=1,2}(\check u_a^\mu \boldsymbol{J}_{\mu\nu} b^\nu)u_{a}^{[\alpha} b^{\beta]}/b^2 + (\check u_1^\mu \boldsymbol{J}_{\mu\nu} \check u_2^\nu)u_{1}^{[\alpha} u_{2}^{\beta]}\,.
\end{equation}
Finally, inserting the waveform kernel~\eqref{eq:waveformkernel} into
Eqs.~\eqref{eq:energymomentumWW} and \eqref{eq:angularmomentumWW}
we find 
\begin{subequations}\label{eq:PJexpandedout}
\begin{align}
\label{eq:Pexpandedout}
P^\alpha &= \int_k k^\alpha 
(\tilde{\mathcal{A}}_0^\ast \tilde{\mathcal{A}}_0 
+ 
2\operatorname{Re}[\tilde{\mathcal{A}}^\ast_0 \tilde{\mathcal{B}}_1]
-
2\operatorname{Im}[\tilde{\mathcal{A}}^\ast_0 \tilde{\mathcal{C}}_1]
)
\,,
\\
\begin{split} \label{eq:Jexpandedout}
\boldsymbol{J}^{\alpha\beta} 
 &= \int_k \big(-i
 \boldsymbol{O}^{\alpha\beta}[\tilde{\mathcal{A}}_0,\tilde{\mathcal{A}}_0]
 +
 2\operatorname{Im} \boldsymbol{O}^{\alpha\beta}[\tilde{\mathcal{A}}_0,\tilde{\mathcal{B}}_1]
\\
 &\qquad+2\operatorname{Re}  \boldsymbol{O}^{\alpha\beta}[\tilde{\mathcal{A}}_0,\tilde{\mathcal{C}}_1]
\big)\,,
\end{split}
\end{align}
\end{subequations}
up to $\mathcal{O}(G^5)$ terms.
Here we introduced the shorthand
\begin{equation}
    \boldsymbol{O}_{\alpha\beta}[\tilde{X},\tilde{Y}]
    =
    \tilde{X}^\ast k^{\phantom{\mu}}_{[\alpha}\frac{\partial}{\partial k^{\beta]}}\tilde{Y} + 2 \tilde{X}^{\ast\phantom{\mu}}_{\mu[\alpha}\tilde{Y}^\mu_{\beta]}\,.
\end{equation}
Using the fact that the momentum-space building blocks are real, we see that in impact parameter space they satisfy $\tilde{\mathcal{A}}^\ast_0(b_a^\alpha)=\tilde{\mathcal{A}}_0(-b_a^\alpha)$ and similarly for $\tilde{\mathcal{B}}_1$ and $\tilde{\mathcal{C}}_1$. Then, in the case of $P^\alpha$, one can use the properties of the integrand \eqref{eq:Pexpandedout} to show that the $u_{a}^\mu P_\mu$ components are fixed by the first two terms on the right-hand side, involving only $\mathcal{A}_0$ and $\mathcal{B}_1$, while $b^\mu P_\mu$ is given by the cross-term between $\mathcal{A}_0$ and $\mathcal{C}_1$~\cite{Heissenberg:2025ocy}.
Therefore, to get $E_\text{loss}$, one can focus on the former. 
As for $\boldsymbol{J}^{\alpha\beta} $,
in \eqref{eq:Jexpandedout}, one can show that the $u_{a}^\mu \boldsymbol{J}_{\mu\nu} b^\nu$ components depend on $\mathcal{A}_0$ and $\mathcal{B}_1$ only, while only $u_1^\mu \boldsymbol{J}_{\mu\nu} u_2^\nu$ is sensitive to $\mathcal{A}_0$ times $\mathcal{C}_1$ \cite{Heissenberg:2025ocy}.
Hence, when targeting $\boldsymbol{J}_\text{rad}$ the latter is irrelevant. 

Notice also that $u_{a}^\mu \boldsymbol{J}_{\mu\nu} b^\nu$ will only mix with the $u_{a}^\mu P_\mu$ components for any translation \eqref{eq:translationJ} along $b^\alpha$, such as the one needed to move from the frame where particle 2 is at the origin to the one where the center-of-mass is at the origin. Conversely, the $u_1^\mu \boldsymbol{J}_{\mu\nu} u_2^\nu$ mass dipole component transforms nontrivially under translations along $u_{a}^\alpha$, and for this reason it is sensitive to subtle infrared effects plaguing the imaginary part $\mathcal{C}_1$. These include an unspecified time translation in the center-of-mass frame induced by re-scattering effects \cite{Caron-Huot:2023vxl}, transformations under BMS supertranslations \cite{Georgoudis:2023eke} connecting different values of the initial shear \cite{Veneziano:2022zwh}, and a spatial translation along the particles' velocities \cite{Bini:2026dvn,Bini:2026jwc}. For these reasons, we will not evaluate the last term in \eqref{eq:Jdecomp}, which also drops when considering the dual angular momentum vector in the center-of-mass frame,\footnote{In that frame, the initial mechanical angular momentum is given by $J_\text{mech}^{\alpha\beta}=b^{[\alpha}p^{\beta]}$ or $J_\text{mech}^\alpha=-\frac{1}{2}\varepsilon^{\alpha\beta\rho\sigma}t_\beta (J_{\text{mech}})_{\rho\sigma}$. By the conservation law, its variation due to dissipative effects must obey $\delta J_\text{mech}^{\alpha\beta} = -J^{\alpha\beta}$. Since in the nonspinning case $\delta t^\alpha=a_b b^\alpha + a_p p^\alpha$, it follows that $\delta J_\text{mech}^\alpha=-J^{\alpha}$.}
$J^\mu = -\tfrac{1}{2} \varepsilon^{\mu\nu\rho\sigma} t_{\nu} J_{\rho\sigma}
$,
with $\varepsilon_{0123}=1$, and the mass dipole component $K_\text{loss}=\tfrac{1}{b}b^\mu J_{\mu\nu} t^\nu$.
Aligning $b^\mu$ with the $x$-axis and $p^\mu$ with the $y$-axis, $J^z=J_{xy}=J_\text{loss}$ and $J_{x0}=K_\text{loss}$ are our key observables.

 \paragraph*{\textbf{Integrand Construction.}}
We thus target the terms of \Eqn{eq:PJexpandedout} that
contribute to $E_\text{loss}$, $\boldsymbol{J}_\text{rad}$ at $\mathcal{O}(G^4)$:
\begin{equation}\label{eq:relevantcomponents}
    P_\parallel^\alpha = 2\operatorname{Re}\!\! \int_k\!\! k^\alpha \tilde{\mathcal{A}}^\ast_{0} \tilde{\mathcal{B}}^{\phantom{\ast}}_1\,,
    \quad
    \boldsymbol{J}_\perp^{\alpha\beta} = 2\operatorname{Im} \!\!\int_k\!\! \boldsymbol{O}^{\alpha\beta}[\tilde{\mathcal{A}}_{0},\tilde{\mathcal{B}}^{\phantom{\ast}\!\!}_1]\,.
\end{equation}
We exploit the technique of reverse unitarity \cite{Anastasiou:2002qz,Anastasiou:2002qz,Anastasiou:2003yy,Anastasiou:2015yha},
adapted to gravitational-wave applications in \Rcites{Herrmann:2021lqe,DiVecchia:2022piu}.
The idea is to recast the expressions \eqref{eq:relevantcomponents},
which involve integrals of products of Fourier transforms, as convolutions
taking the form of cut three-loop integrals. 

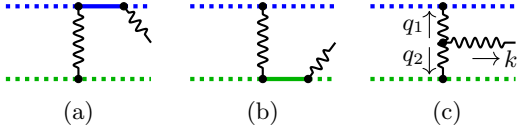
\begin{figure}[t!]
		\centering
		\begin{subfigure}{0.13\textwidth}
			\centering
			\begin{tikzpicture}[baseline={(current bounding box.center)},scale=.6]
				\coordinate (inA) at (-.6,.8);
				\coordinate (outA) at (2.6,.8);
				\coordinate (inB) at (-.6,-.8);
				\coordinate (outB) at (2.6,-.8);
				\coordinate (xA) at (0,.8);
				\coordinate (yA) at (1,.8);
				\coordinate (zA) at (2,.8);
				\coordinate (xB) at (0,-.8);
				\coordinate (yB) at (1,-.8);
				\coordinate (k) at (2.6,0);
				\draw [dotted,ultra thick,blue] (inA) -- (yA);
				\draw [ultra thick,blue] (yA) -- (zA);
				\draw [dotted,ultra thick,blue] (zA) -- (outA);
				\draw [dotted,ultra thick,green!70!black] (inB) -- (outB);
				\draw [fill] (yA) circle (.08);
				\draw [fill] (yB) circle (.08);
				\draw [fill] (zA) circle (.08);
				\draw [photon] (yA) -- (yB);
				\draw [photon] (zA) -- (k);
			\end{tikzpicture}
			\caption{}
		\end{subfigure}
		\begin{subfigure}{0.13\textwidth}
			\centering
			\begin{tikzpicture}[baseline={(current bounding box.center)},scale=.6]
				\coordinate (inA) at (-.6,.8);
				\coordinate (outA) at (2.6,.8);
				\coordinate (inB) at (-.6,-.8);
				\coordinate (outB) at (2.6,-.8);
				\coordinate (xA) at (0,.8);
				\coordinate (yA) at (1,.8);
				\coordinate (xB) at (0,-.8);
				\coordinate (yB) at (1,-.8);
				\coordinate (zB) at (2,-.8);
				\coordinate (k) at (2.6,0);
				\draw [dotted,ultra thick,blue] (inA) -- (yA);
				\draw [dotted,ultra thick,blue] (yA) -- (outA);
				\draw [dotted,ultra thick,green!70!black] (inB) -- (yB);
				\draw [dotted,ultra thick,green!70!black] (zB) -- (outB);
				\draw [ultra thick,green!70!black] (yB) -- (zB);
				\draw [fill] (yA) circle (.08);
				\draw [fill] (yB) circle (.08);
				\draw [fill] (zB) circle (.08);
				\draw [photon] (yA) -- (yB);
				\draw [photon] (zB) -- (k);
			\end{tikzpicture}
			\caption{}
		\end{subfigure}
        \begin{subfigure}{0.13\textwidth}
			\centering
			\begin{tikzpicture}[baseline={(current bounding box.center)},scale=.6]
				\coordinate (inA) at (-.6,.8);
				\coordinate (outA) at (2.6,.8);
				\coordinate (inB) at (-.6,-.8);
				\coordinate (outB) at (2.6,-.8);
				\coordinate (xA) at (0,.8);
				\coordinate (yA) at (1,.8);
				\coordinate (xB) at (0,-.8);
				\coordinate (yB) at (1,-.8);
				\coordinate (yC) at (1,0);
				\coordinate (k) at (2.6,0);
				\draw [dotted,ultra thick,blue] (inA) -- (yA);
				\draw [dotted,ultra thick,blue] (yA) -- (outA);
				\draw [dotted,ultra thick,green!70!black] (inB) -- (outB);
				\draw [fill] (yA) circle (.08);
				\draw [fill] (yB) circle (.08);
				\draw [fill] (yC) circle (.08);
				\draw [photon] (yA) -- (yB);
				\draw [photon] (yC) -- (k);
				\node at (2.5,-.35) {$k$};
				\draw [thin,->] (1.65,-.35) -- (2.25,-.35);
				\node at (.35,-.35) {$q_2$};
				\draw [thin,->] (.7,.1) -- (.7,.7);
				\node at (.35,.35) {$q_1$};
				\draw [thin,->] (.7,-.1) -- (.7,-.7);
				\end{tikzpicture}
			\caption{}
		\end{subfigure}
		\caption{\small Worldline diagrams for the tree-level waveform.
		}
		\label{fig:lradiation}
	\end{figure}

\begin{figure*}[ht!]
		\centering
        \begin{subfigure}{0.13\textwidth}
			\centering
			\begin{tikzpicture}[baseline={(current bounding box.center)},scale=.6]
				\coordinate (inA) at (-.6,.8);
				\coordinate (outA) at (2.6,.8);
				\coordinate (inB) at (-.6,-.8);
				\coordinate (outB) at (2.6,-.8);
				\coordinate (xA) at (0,.8);
				\coordinate (yA) at (1,.8);
				\coordinate (zA) at (2,.8);
				\coordinate (xB) at (0,-.8);
				\coordinate (yB) at (1,-.8);
				\coordinate (k) at (2.6,0);
				\draw [dotted,ultra thick,blue] (inA) -- (xA);
				\draw [ultra thick,blue] (xA) -- (yA);
				\draw [ultra thick,blue] (yA) -- (zA);
				\draw [dotted,ultra thick,blue] (zA) -- (outA);
				\draw [dotted,ultra thick,green!70!black] (inB) -- (outB);
				\draw [fill] (xA) circle (.08);
				\draw [fill] (xB) circle (.08);
				\draw [fill] (yA) circle (.08);
				\draw [fill] (yB) circle (.08);
				\draw [fill] (zA) circle (.08);
				\draw [photon] (xA) -- (xB);
				\draw [photon] (yA) -- (yB);
				\draw [photon] (zA) -- (k);
				\node at (-.35,0) {$\ell$};
				\draw[thin,->] (.25,0) arc[start angle=180, end angle=-115, radius=.25];
			\end{tikzpicture}
			\caption{}
            \label{fig:top1}
		\end{subfigure}
        \begin{subfigure}{0.13\textwidth}
			\centering
			\begin{tikzpicture}[baseline={(current bounding box.center)},scale=.6]
				\coordinate (inA) at (-.6,.8);
				\coordinate (outA) at (2.6,.8);
				\coordinate (inB) at (-.6,-.8);
				\coordinate (outB) at (2.6,-.8);
				\coordinate (xA) at (0,.8);
				\coordinate (yA) at (1,.8);
				\coordinate (xB) at (0,-.8);
				\coordinate (yB) at (1,-.8);
				\coordinate (yC) at (1,0);
				\coordinate (k) at (2.6,0);
				\draw [dotted,ultra thick,blue] (inA) -- (xA);
				\draw [dotted,ultra thick,blue] (yA) -- (outA);
				\draw [dotted,ultra thick,green!70!black] (inB) -- (outB);
				\draw [ultra thick,blue] (xA) -- (yA);
				\draw [fill] (xA) circle (.08);
				\draw [fill] (xB) circle (.08);
				\draw [fill] (yA) circle (.08);
				\draw [fill] (yB) circle (.08);
				\draw [fill] (yC) circle (.08);
				\draw [photon] (xA) -- (xB);
				\draw [photon] (yA) -- (yB);
				\draw [photon] (yC) -- (k);
			\end{tikzpicture}
			\caption{}
            \label{fig:top2}
		\end{subfigure}
        \begin{subfigure}{0.13\textwidth}
			\centering
			\begin{tikzpicture}[baseline={(current bounding box.center)},scale=.6]
				\coordinate (inA) at (-.6,.8);
				\coordinate (outA) at (2.6,.8);
				\coordinate (inB) at (-.6,-.8);
				\coordinate (outB) at (2.6,-.8);
				\coordinate (xA) at (0,.8);
				\coordinate (yA) at (1,.8);
				\coordinate (xB) at (0,-.8);
				\coordinate (yB) at (1,-.8);
				\coordinate (zB) at (2,-.8);
				\coordinate (k) at (2.6,0);
				\draw [dotted,ultra thick,blue] (inA) -- (xA);
				\draw [dotted,ultra thick,blue] (yA) -- (outA);
				\draw [dotted,ultra thick,green!70!black] (inB) -- (yB);
				\draw [dotted,ultra thick,green!70!black] (zB) -- (outB);
				\draw [ultra thick,blue] (xA) -- (yA);
				\draw [ultra thick,green!70!black] (yB) -- (zB);
				\draw [fill] (xA) circle (.08);
				\draw [fill] (xB) circle (.08);
				\draw [fill] (yA) circle (.08);
				\draw [fill] (yB) circle (.08);
				\draw [fill] (zB) circle (.08);
				\draw [photon] (xA) -- (xB);
				\draw [photon] (yA) -- (yB);
				\draw [photon] (zB) -- (k);
			\end{tikzpicture}
			\caption{}
            \label{fig:top3}
		\end{subfigure}
		\begin{subfigure}{0.13\textwidth}
			\centering
			\begin{tikzpicture}[baseline={(current bounding box.center)},scale=.6]
				\coordinate (inA) at (-.6,.8);
				\coordinate (outA) at (2.6,.8);
				\coordinate (inB) at (-.6,-.8);
				\coordinate (outB) at (2.6,-.8);
				\coordinate (xA) at (0,.8);
				\coordinate (yA) at (1,.8);
				\coordinate (zA) at (2,.8);
				\coordinate (xB) at (0,-.8);
				\coordinate (zB) at (2,-.8);
				\coordinate (zC) at (2,0);
				\coordinate (k) at (2.8,0);
				\draw [dotted,ultra thick,blue] (inA) -- (xA);
				\draw [ultra thick,blue] (xA) -- (yA);
				\draw [ultra thick,blue] (yA) -- (zA);
				\draw [dotted,ultra thick,blue] (zA) -- (outA);
				\draw [dotted,ultra thick,green!70!black] (inB) -- (outB);
				\draw [fill] (xA) circle (.08);
				\draw [fill] (xB) circle (.08);
				\draw [fill] (yA) circle (.08);
				\draw [fill] (zA) circle (.08);
				\draw [fill] (zB) circle (.08);
				\draw [photon] (xA) -- (xB);
				\draw [photon] (zA) -- (zB);
				\draw[fill=white, draw=none] (zC) circle (0.3);
				\draw [photon] (yA) to[out=-90,in=0] (k);
			\end{tikzpicture}
			\caption{}
            \label{fig:top4}
		\end{subfigure}
        \begin{subfigure}{0.13\textwidth}
			\centering
			\begin{tikzpicture}[baseline={(current bounding box.center)},scale=.6]
				\coordinate (inA) at (-.6,.8);
				\coordinate (outA) at (2.6,.8);
				\coordinate (inB) at (-.6,-.8);
				\coordinate (outB) at (2.6,-.8);
				\coordinate (xyA) at (0.5,.8);
				\coordinate (xyC) at (0.5,0);
				\coordinate (zA) at (2,.8);
				\coordinate (xB) at (0,-.8);
				\coordinate (yB) at (1,-.8);
				\coordinate (k) at (2.6,0);
				\draw [dotted,ultra thick,blue] (inA) -- (xyA);
				\draw [ultra thick,blue] (xyA) -- (zA);
				\draw [dotted,ultra thick,blue] (zA) -- (outA);
				\draw [dotted,ultra thick,green!70!black] (inB) -- (outB);
				\draw [fill] (xyA) circle (.08);
				\draw [fill] (xyC) circle (.08);
				\draw [fill] (xB) circle (.08);
				\draw [fill] (yB) circle (.08);
				\draw [fill] (zA) circle (.08);
				\draw [photon] (xyC) -- (xB);
				\draw [photon] (xyC) -- (yB);
				\draw [photon] (xyA) -- (xyC);
				\draw [photon] (zA) -- (k);
			\end{tikzpicture}
			\caption{}
            \label{fig:top5}
		\end{subfigure}
		
		\vspace{1pt}
		
        \begin{subfigure}{0.13\textwidth}
			\centering
			\begin{tikzpicture}[baseline={(current bounding box.center)},scale=.6]
				\coordinate (inA) at (-1,.8);
				\coordinate (outA) at (2,.8);
				\coordinate (inB) at (-1,-.8);
				\coordinate (outB) at (2,-.8);
				\coordinate (C) at (.5,-.2);
				\coordinate (D) at (.5,.2);
				\coordinate (A) at (.5,.8);
				\coordinate (xB) at (-.4,-.8);
				\coordinate (yB) at (1.4,-.8);
				\coordinate (k) at (2,.2);
				\draw [dotted,ultra thick,blue] (inA) -- (A);
				\draw [dotted,ultra thick,blue] (A) -- (outA);
				\draw [dotted,ultra thick,green!70!black] (inB) -- (outB);
				\draw [fill] (C) circle (.08);
				\draw [fill] (D) circle (.08);
				\draw [fill] (A) circle (.08);
				\draw [fill] (xB) circle (.08);
				\draw [fill] (yB) circle (.08);
				\draw [photon] (C) -- (A);
				\draw [photon] (xB) -- (C);
				\draw [photon] (yB) -- (C);
				\draw [photon] (D) -- (k);
			\end{tikzpicture}
			\caption{}
            \label{fig:top6}
		\end{subfigure}
        \begin{subfigure}{0.13\textwidth}
			\centering
			\begin{tikzpicture}[baseline={(current bounding box.center)},scale=.6]
				\coordinate (inA) at (-1,.8);
				\coordinate (outA) at (2,.8);
				\coordinate (inB) at (-1,-.8);
				\coordinate (outB) at (2,-.8);
				\coordinate (C) at (.5,0);
				\coordinate (A) at (.5,.8);
				\coordinate (xB) at (-.4,-.8);
				\coordinate (yB) at (1.4,-.8);
				\coordinate (k) at (2,0);
				\draw [dotted,ultra thick,blue] (inA) -- (A);
				\draw [dotted,ultra thick,blue] (A) -- (outA);
				\draw [dotted,ultra thick,green!70!black] (inB) -- (outB);
				\draw [fill] (C) circle (.08);
				\draw [fill] (A) circle (.08);
				\draw [fill] (xB) circle (.08);
				\draw [fill] (yB) circle (.08);
				\draw [photon] (C) -- (A);
				\draw [photon] (xB) -- (C);
				\draw [photon] (yB) -- (C);
				\draw [photon] (C) -- (k);
			\end{tikzpicture}
			\caption{}
            \label{fig:top6bis}
		\end{subfigure}
        \begin{subfigure}{0.13\textwidth}
			\centering
			\begin{tikzpicture}[baseline={(current bounding box.center)},scale=.6]
				\coordinate (inA) at (-1,.8);
				\coordinate (outA) at (2,.8);
				\coordinate (inB) at (-1,-.8);
				\coordinate (outB) at (2,-.8);
				\coordinate (C) at (.9,-.3);
				\coordinate (D) at (.5,.2);
				\coordinate (A) at (.5,.8);
				\coordinate (xB) at (-.4,-.8);
				\coordinate (yB) at (1.4,-.8);
				\coordinate (k) at (2,.2);
				\draw [dotted,ultra thick,blue] (inA) -- (A);
				\draw [dotted,ultra thick,blue] (A) -- (outA);
				\draw [dotted,ultra thick,green!70!black] (inB) -- (outB);
				\draw [fill] (C) circle (.08);
				\draw [fill] (D) circle (.08);
				\draw [fill] (A) circle (.08);
				\draw [fill] (xB) circle (.08);
				\draw [fill] (yB) circle (.08);
				\draw [photon] (D) -- (A);
				\draw [photon] (xB) -- (D);
				\draw [photon] (yB) -- (D);
				\draw [photon] (C) -- (k);
			\end{tikzpicture}
			\caption{}
            \label{fig:top7}
		\end{subfigure}
        \begin{subfigure}{0.13\textwidth}
			\centering
			\begin{tikzpicture}[baseline={(current bounding box.center)},scale=.6]
				\coordinate (inA) at (-.6,.8);
				\coordinate (outA) at (2.6,.8);
				\coordinate (inB) at (-.6,-.8);
				\coordinate (outB) at (2.6,-.8);
				\coordinate (yB) at (0.5,.8);
				\coordinate (xyC) at (0.5,0);
				\coordinate (zA) at (2,-.8);
				\coordinate (xB) at (0,-.8);
				\coordinate (xyA) at (1,-.8);
				\coordinate (k) at (2.6,0);
				\draw [dotted,ultra thick,blue] (inA) -- (outA);
				\draw [ultra thick,green!70!black] (xyA) -- (zA);
				\draw [dotted,ultra thick,green!70!black] (inB) -- (outB);
				\draw [fill] (xyA) circle (.08);
				\draw [fill] (xyC) circle (.08);
				\draw [fill] (xB) circle (.08);
				\draw [fill] (yB) circle (.08);
				\draw [fill] (zA) circle (.08);
				\draw [photon] (xyC) -- (xB);
				\draw [photon] (xyC) -- (yB);
				\draw [photon] (xyA) -- (xyC);
				\draw [photon] (zA) -- (k);
			\end{tikzpicture}
			\caption{}
            \label{fig:top8}
		\end{subfigure}
        \begin{subfigure}{0.13\textwidth}
			\centering
			\begin{tikzpicture}[baseline={(current bounding box.center)},scale=.6]
				\coordinate (inA) at (-.6,.8);
				\coordinate (outA) at (2.6,.8);
				\coordinate (inB) at (-.6,-.8);
				\coordinate (outB) at (2.6,-.8);
				\coordinate (xA) at (0,.8);
				\coordinate (yA) at (1,.8);
				\coordinate (xB) at (0,-.8);
				\coordinate (yB) at (1,-.8);
				\coordinate (zB) at (2,-.8);
				\coordinate (k) at (2.6,0);
                \coordinate (tV) at (1.5,0);
				\draw [dotted,ultra thick,blue] (inA) -- (outA);
				\draw [dotted,ultra thick,green!70!black] (inB) -- (outB);
				\draw [ultra thick,green!70!black] (xB) -- (yB);
				\draw [fill] (xA) circle (.08);
				\draw [fill] (xB) circle (.08);
				\draw [fill] (yB) circle (.08);
				\draw [fill] (zB) circle (.08);
                \draw [fill] (tV) circle (.08);
				\draw [photon] (xA) -- (xB);
                \draw [photon] (zB) -- (tV);
                \draw [photon] (yB) -- (tV);
				\draw [photon] (tV) -- (k);
			\end{tikzpicture}
			\caption{}
            \label{fig:top9}
		\end{subfigure}
        \begin{subfigure}{0.13\textwidth}
			\centering
			\begin{tikzpicture}[baseline={(current bounding box.center)},scale=.6]
				\coordinate (inA) at (-.6,.8);
				\coordinate (outA) at (2.6,.8);
				\coordinate (inB) at (-.6,-.8);
				\coordinate (outB) at (2.6,-.8);
				\coordinate (xA) at (0,.8);
				\coordinate (yA) at (1,.8);
				\coordinate (xB) at (0,-.8);
				\coordinate (yB) at (.5,-.8);
				\coordinate (zB) at (1.5,-.8);
                \coordinate (tB) at (2,-.8);
				\coordinate (k) at (2.6,0);
                \coordinate (tV) at (1,0);
				\draw [dotted,ultra thick,blue] (inA) -- (outA);
				\draw [dotted,ultra thick,green!70!black] (inB) -- (outB);
				\draw [ultra thick,green!70!black] (xB) -- (yB);
                \draw [ultra thick,green!70!black] (zB) -- (tB);
				\draw [fill] (xA) circle (.08);
				\draw [fill] (xB) circle (.08);
				\draw [fill] (yB) circle (.08);
				\draw [fill] (zB) circle (.08);
                \draw [fill] (tB) circle (.08);
				\draw [photon] (xA) -- (xB);
                \draw[photon] (yB) arc[start angle=180, end angle=0, radius=.5];
				\draw [photon] (tB) -- (k);
			\end{tikzpicture}
			\caption{}
            \label{fig:top10}
		\end{subfigure}
		\caption{\small Worldline Feynman diagrams that determine the one-loop waveform integrand $\mathcal{N}_1^{\mu\nu}$.}
		\label{fig:subradiationGraphs}
	\end{figure*}
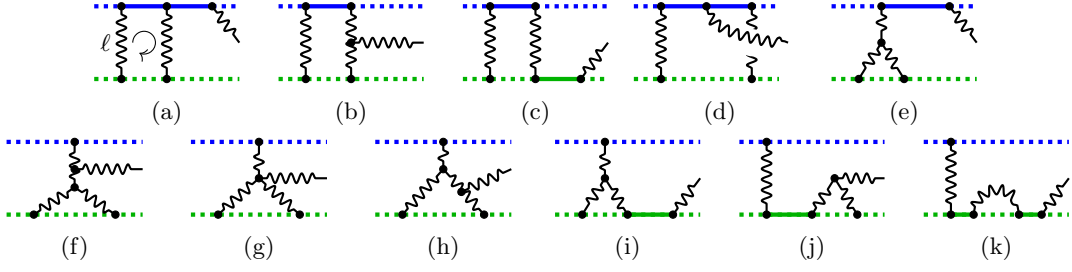

The first ingredient is the tree $2\to3$ amplitude \cite{Luna:2017dtq} depicted in Fig.~\ref{fig:lradiation} (see \Eqn{eq:treelevel} for the full expression). 
	In agreement with the Ward Identity \eqref{eq:WardIdentity}, $k_\mu \mathcal{A}_0^{\mu\nu}=0$.
The second ingredient is a simple integrand $\mathcal{N}_1$ for the one-loop waveform, which we obtain from the worldline approach \cite{Mogull:2020sak} (see also \cite{Brunello:2025eso,He:2025how}). While $\mathcal{A}_0$ is a monomial proportional to $m_1^2 m_2^2$, $\mathcal{N}_1$ has two terms, proportional to $m_1^3 m_2^2$ and $m_1^2 m_2^3$, which are mapped to each other by particle-interchange symmetry $1\leftrightarrow2$. We shall focus on the contribution with the larger power of $m_2$, which is obtained from the diagrams in Fig.~\ref{fig:subradiationGraphs}, and restore $1\leftrightarrow2$ symmetry at the end.
Crucially, by construction, this integrand always involves a cut on the heavy line, $\delta(u_2\cdot\ell)$. 
We checked that $k_\mu \mathcal{N}_1^{\mu\nu}$ vanishes upon the integration over the loop momentum, again in agreement with \Eqn{eq:WardIdentity}. The real part $\mathcal{B}_1$ can then be conveniently selected by taking the Principal Value (PV), so that 
\begin{equation}\label{eq:connectBN}
    \mathcal{B}_{1}^{\mu\nu}(q_1,k)= \int \frac{d^D\ell}{(2\pi)^{D}}\,\operatorname{PV}\mathcal{N}^{\mu\nu}_{1}(q_1,k,\ell)\,.
\end{equation}
We provide $\mathcal{A}_0$ and $\mathcal{N}_1$ in the file \texttt{anc-integrands.m}.

In this way, we recast \Eqn{eq:relevantcomponents} as the Fourier transform \eqref{eq:elasticFT}, $P_\parallel^\alpha=\operatorname{FT}\mathbb{P}_\parallel^\alpha$, of the momentum-space kernel
\begin{equation}\label{eq:revunitP}
    \tfrac{1}{2}\mathbb{P}_\parallel^\alpha
    =
    \int
    k^{\alpha}
    \mathcal{N}_1^{L} \mathcal{A}_0^R
    d\mu\,,
\end{equation}
where $\mathcal{N}_1^L =\operatorname{PV} \mathcal{N}_1(q_1,k,\ell)$, $\mathcal{A}_0^R=\mathcal{A}_0(q-q_1,k+q_1-q)$
and the measure $d\mu$ is given by
\begin{equation}
    d\mu =\theta(k^0)\delta(k^2)\delta(2p_1\cdot q_1)\delta(2p_2\cdot(q_1+ k))\frac{d^Dq_1 \,d^D\ell\, d^Dk}{(2\pi)^{3(D-1)}}\,.
\end{equation}
In \Eqn{eq:revunitP} we thus achieved the goal of casting the integral as the cut illustrated in Fig.~\ref{fig:threePC3loop}.
Similarly, in the translation frame $b^\alpha_2=0$, we obtain
$\boldsymbol{J}_\perp^{\alpha\beta}=\operatorname{FT}\mathbb{J}_\perp^{\alpha\beta}$ with
\begin{align}\label{eq:revunitJ}
    &\tfrac{1}{2}\mathbb{J}_{\perp \alpha\beta}
    =
    \int
    k^{\phantom{\mu}}_{[\alpha}\frac{\partial}{\partial k^{\beta]}}
    \left[\mathcal{N}^{L}_1 d\mu\right]
    \mathcal{A}_0^R
    \\
    &
    -u_{2[\alpha} 
    \int
    k_{\beta]}
    \frac{\partial}{\partial q_{\parallel2}}\left[    
    \mathcal{N}^{L}_1 \mathcal{A}_0^R
    \right] d\mu
    -2
    \int
    \left(\mathcal{N}^{L}_1\right)_{\mu[\alpha}\left(\mathcal{A}_0^{R}\right)^\mu_{\beta]}
    d\mu,\nonumber
\end{align}
where $q_{\parallel2}=-u_2\cdot q$. 
While $\mathbb{P}_\parallel^\alpha$~\eqref{eq:revunitP} only involves contractions between the two ingredients,
$\mathbb{J}_{\perp \alpha\beta}$~\eqref{eq:revunitJ} also contains derivatives with respect to $k^\mu$, which can act both on $\mathcal{N}_1$ and on the measure $d\mu$ in its first line,
and to $q_{\parallel 2}$ in the second line.
Nevertheless, $\mathbb{J}_{\perp \alpha\beta}$ also achieves the form in Fig.~\ref{fig:threePC3loop}.

We remark that the nonplanar topology in Fig.~\ref{fig:top4} can be planarized by applying the partial-fractioning identity
\begin{equation}
    \frac{1}{z^2 (z+\omega )^2} =
	\frac{\omega -2 z}{z^2 \omega ^3}+\frac{2 z+3 \omega }{\omega ^3 (z+\omega )^2}\,,
\end{equation}
with $z=u_1\cdot \ell$ and $\omega=u_1\cdot q_2$, and by subsequently relabeling $\ell\mapsto -\ell-q_2$ in the second term. This effectively maps such a nonplanar topology to the one in Fig.~\ref{fig:top1}. 

\paragraph*{\textbf{Integration.}}
We then organize the calculation of the integrals in \eqref{eq:revunitP}, \eqref{eq:revunitJ} in terms of the following \emph{planar} family by identifying $q_1=\ell_3$,  $k=\ell_2-\ell_3$, $\ell=\ell_1$, 
\begin{align}\label{eq:Jfamily}
    I^{m_1m_2m_3}_{n_1\cdots n_{12}} &= \int \frac{\delta^{[m_1]}(u_2\cdot \ell_1)\delta^{[m_2]}(u_2\cdot \ell_2)\delta^{[m_3]}(u_1\cdot\ell_3)}{D_1^{n_1}D_2^{n_2} D_3^{n_3} D_{4}^{n_4}}
    \nonumber\\
    &\times \frac{2\pi\theta(\ell_2^0-\ell_3^0)\delta((\ell_2-\ell_3)^2)}{D_{6}^{n_6} D_7^{n_7}\cdots D_{12}^{n_{12}}}\prod_{i=1}^3 \frac{d^D\ell_i}{\pi^{D/2}}\,,
\end{align}
where $\delta^{[m+1]}(x) = 2\pi(-1)^{m} (m!)^{-1} \delta^{(m)}(2x)$ stands for (derivatives of) the delta function, 
\begin{equation}
    D_1 = 2u_1\cdot\ell_1\,,\ 
    D_2 = 2u_1\cdot\ell_2\,,\ 
    D_3 = 2u_2\cdot\ell_3
\end{equation}
and
\begin{align}
\nonumber
D_4&=(\ell_1-\ell_3)^2\,,\ 
D_6=(\ell_1-\ell_2)^2\,,\\ 
D_7&=\ell_1^2\,,\ 
D_8=\ell_2^2\,,\ 
D_9=\ell_3^2\,,\\
D_{10}&=(\ell_1-q)^2\,,\ 
D_{11}=(\ell_2-q)^2\,,\ 
D_{12}=(\ell_3-q)^2\,.
\nonumber 
\end{align}
The first delta function on the right-hand side of \Eqn{eq:Jfamily} is the cut appearing in the one-loop integrand $\mathcal{N}_1$ (the dashed green line in Fig.~\ref{fig:subradiationGraphs}),
while the remaining three are due to the three-particle cut gluing
$\mathcal{N}_1$ and $\mathcal{A}_0$ (Fig.~\ref{fig:threePC3loop}).
Compared with the integrals appearing in \Rcites{Jakobsen:2023ndj,Jakobsen:2023hig},
we thus have an additional cut, localizing $D_5^{-1}=1/(\ell_2-\ell_3)^2$ to the graviton's future light cone.
In our integrand, $D_{10}$ only appears as a numerator, $n_{10}\le0$. Moreover, as already remarked, at most one propagator can approach the mass-shell, and in that case we employ the PV prescription to select the contribution from $\mathcal{B}_{1}$. 

The integrals of the family $I$ \eqref{eq:Jfamily} decompose into an even-in-$q$ sector and an odd-in-$q$ sector, which do not mix under Integration-By-Parts (IBP) identities. Since we focus on the $u_a^\mu P_{\mu}$ and $u_{a}^\mu \boldsymbol{J}_{\mu\nu} b^\nu$ components, which are the only ones entering \Eqn{eq:relevantcomponents}, we are only interested in the even-in-$q$ sector, for which we find 54 master integrals. We solve them by the method of canonical  differential equations \cite{Henn:2013pwa} with respect to the velocity parameter $x=\gamma-\sqrt{\gamma^2-1}$. We employed \texttt{LiteRed} \cite{Lee:2012cn,Lee:2013mka} and \texttt{Kira} \cite{Maierhofer:2017gsa,Klappert:2020nbg,Lange:2025fba} to implement the IBP reduction.

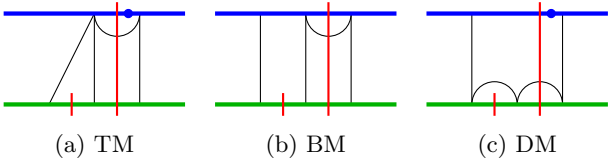
\begin{figure}[t]
	\begin{subfigure}{0.15\textwidth}
		\centering
		\begin{tikzpicture}[scale=.6]
			\draw[thin] (-1,0)--(0,2);
			\draw[thin] (0,2) arc[start angle=180, end angle=360, radius=.5];
			\draw[thin] (0,0)--(0,2);
			\draw[thin] (1,0)--(1,2);
			\draw [blue, fill] (.75,2) circle (.09);
			\draw[ultra thick,blue] (-2, 2) -- (2,2);
			\draw[ultra thick,green!70!black] (-2, 0) -- (2, 0);
			\draw[thick,red] (-.5,-.25)--(-.5,.25);
			\draw[thick,red] (.5,-.25)--(.5,2.25);
		\end{tikzpicture}
		\caption{TM}
		\label{fig:TM}
	\end{subfigure}
	\begin{subfigure}{0.15\textwidth}
		\centering
		\begin{tikzpicture}[scale=.6]
			\draw[thin] (-1,0)--(-1,2);
			\draw[thin] (0,2) arc[start angle=180, end angle=360, radius=.5];
			\draw[thin] (0,0)--(0,2);
			\draw[thin] (1,0)--(1,2);
			\draw[ultra thick,blue] (-2, 2) -- (2,2);
			\draw[ultra thick,green!70!black] (-2, 0) -- (2, 0);
			\draw[thick,red] (-.5,-.25)--(-.5,.25);
			\draw[thick,red] (.5,-.25)--(.5,2.25);
		\end{tikzpicture}
		\caption{BM}
		\label{fig:BM}
	\end{subfigure}
	\begin{subfigure}{0.15\textwidth}
		\centering
		\begin{tikzpicture}[scale=.6]
			\draw[thin] (-1,0)--(-1,2);
			\draw[thin] (-1,0) arc[start angle=180, end angle=0, radius=.5];
			\draw[thin] (0,0) arc[start angle=180, end angle=0, radius=.5];
			\draw[thin] (1,0)--(1,2);
			\draw[ultra thick,blue] (-2, 2) -- (2,2);
			\draw[ultra thick,green!70!black] (-2, 0) -- (2, 0);
			\draw[thick,red] (-.5,-.25)--(-.5,.25);
			\draw[thick,red] (.5,-.25)--(.5,2.25);
			\draw [blue, fill] (.75,2) circle (.09);
		\end{tikzpicture}
		\caption{DM}
		\label{fig:MM}
	\end{subfigure}
	\caption{Integrals relevant for the boundary conditions. A dot means $m_3=2$ in \eqref{eq:Jfamily} and thus a $\delta'$ distribution.}
	\label{fig:BCintegrals}
\end{figure}

To fix the boundary conditions in the small-velocity limit,
$x\to1$ or $p_\infty=\sqrt{\gamma^2-1}\to0$, we exploit the method of regions.
Regions are characterized by the number of gravitons that
can go on-shell (radiative or ``rad''), or stay off-shell (potential).
The 1rad integrals can be IBP-reduced to the Triangle-Mushroom (TM)
and Box-Mushroom (BM) integrals,
and 2rad integrals to Double-Mushroom (DM) --- see Fig.~\ref{fig:BCintegrals}.
In particular,
\begin{equation}
    \text{TM}=\frac{4^{\epsilon} \pi^{\frac{5}{2}} p_{\infty}^{-2\epsilon} \Gamma\left(\frac{1}{2} - 2\epsilon\right) \Gamma\left(\frac{1}{2} - \epsilon\right) \Gamma(\epsilon)}{\Gamma\left(\frac{3}{2} - 5\epsilon\right) \left(q^2\right)^{3\epsilon}} \frac{\sin \pi \epsilon}{\sin 3\pi \epsilon}\,,
\end{equation}
and $\mathrm{BM}=(-\tfrac{1}{2}+{\mathcal O}(\epsilon^2)) \mathrm{TM}$.\footnote{\unexpanded{This can be checked by using the following integral representation
\[
\begin{aligned}
\mathrm{BM}={}&
-\frac{2\pi^{7/2}\Gamma(1-3\epsilon)}
{\sin(3\pi\epsilon)\Gamma(1-\epsilon)
 \Gamma(\tfrac32-5\epsilon)}
\frac{p_\infty^{-2\epsilon}}{(q^2)^{3\epsilon}}
\\
&\times\int_0^\infty dt\int_0^1 du\,
\frac{u^{2-10\epsilon}}
{t^{2\epsilon}(1+t^2)^{2\epsilon}
 (1+t^2u^2)^{1-4\epsilon}}
\\
&\times{}_2F_1\!\left(
-2\epsilon,2\epsilon;1-\epsilon;
\frac{1}{1+t^2}
\right).
\end{aligned}
\]}}

The 1rad and 2rad regions are also identified by their parity in $p_\infty$:
the former is \emph{even} and the latter is \emph{odd}.
The 2rad contributions to \Eqn{eq:relevantcomponents} were also calculated in \Rcite{Heissenberg:2025ocy}
by noting that they can be recast as simpler two-loop calculations.
So we focus on 1rad by dropping the Double-Mushroom in the boundary conditions,
and adding back 2rad at the end using the result of \cite{Heissenberg:2025ocy}.

After calculating in this way the contribution proportional to the higher power of $m_2$, we move to the midpoint frame, $b_1^\alpha+b_2^\alpha=0$, and deduce the other mass structure by $1\leftrightarrow2$ symmetry. This interchanges $u_{1}^\alpha\leftrightarrow u_2^\alpha$ and $b_{1}^\alpha\leftrightarrow b_2^\alpha$, and thus changes the sign of $b^\alpha=b_1^\alpha-b_2^\alpha$.
Our $\mathcal{O}(G^4)$ results take the form
\begin{subequations}\label{eq:expl}
\begin{align}
    P_\parallel^\alpha &= \frac{G^4m_1^2m_2^2}{b^4 p_\infty^5} [
     (m_2\mathcal{X} + m_1\mathcal{Y})u_1^\alpha+ 
    1\leftrightarrow2
    ]\,,
    \\
    \boldsymbol{J}_\perp^{\alpha\beta} &= \frac{G^4m_1^2m_2^2}{b^4 p_\infty^5} 
    [
    (m_2\mathcal{Z} + m_1\mathcal{T}) u_1^{[\alpha}b^{\beta]} + 
    1\leftrightarrow2
    ]\,.
\end{align}
\end{subequations}
Here $\mathcal{X},\mathcal{Y},\mathcal{Z},\mathcal{T}$ are linear combinations of the functions $\Pi_a=\pi_a/p_\infty$ with $a=0,\ldots,3$, and $\Pi_0^2$, where 
\begin{align}
\begin{aligned}
    \pi_0&=\operatorname{arccosh}\gamma
    \label{eq:functions}\,,
    \\
    \pi_1&=-\operatorname{Li}_2(1-x)-\frac{1}{4}\pi_0^2\,,
    \\
    \pi_2&=\operatorname{Li}_2(-x)+\frac{1}{4}\pi_0^2-\frac{1}{2}\log(2(\gamma+1))\pi_0+\frac{\pi^2}{12}\,,
    \\
    \pi_3&=\operatorname{Li}_2(-x^2)+\pi_0^2-2\log(2\gamma)\pi_0+\frac{\pi^2}{12}\,,
\end{aligned}
\end{align}
with coefficients that are rational functions of $\gamma$, for the 1rad part, while they involve an extra overall $p_\infty$ for the 2rad part. See Appendix~\ref{app:XYZT} for the explicit expressions. 
$E_\text{rad}$ agrees with Refs.~\cite{Dlapa:2022lmu,Damgaard:2023ttc,Jakobsen:2023hig},
and our new result for $\boldsymbol{J}_\text{rad}$ satisfies small-velocity checks with Refs.~\cite{Heissenberg:2024umh,Heissenberg:2025fcr}. 
All results are gathered in the ancillary file \texttt{anc-lossesG4.m}, together with the lower-PM results which depend also on $\Pi_{-1}=\log((\gamma+1)/2)$.
The functions $\Pi_a$ are analytic in $p_\infty^2$ for $p_\infty<1$.

\paragraph*{\textbf{Static Angular Momentum Loss.}}
The static component of the angular momentum loss, $\mathcal{J}^{\alpha\beta}$,
arises from the interaction between the static ``Coulombic'' field of the binary,
which is sensitive to BMS supertranslations, and small-frequency waves.
Since the soft theorem completely fixes the leading-order behavior for small frequencies ---
the celebrated memory effect in time domain --- 
as a general function of the classical impulse, it is simpler than the dynamical component.
In order to compare with the available PN data and capture the mechanical loss of angular momentum \cite{Riva:2023xxm}, we adopt the so-called ``intrinsic" gauge~\cite{Veneziano:2022zwh},
in which the angular momentum loss begins at $\mathcal{O}(G^2)$.

The static contribution to $J_{\rm loss}$~\eqref{eq:Jtd} is obtained by taking the $u$-independent term in $\partial_\phi h$ and the gravitational memory arising from the integral of $\partial_u h$. The latter includes terms due to soft gravitons attached to the massive particles. This contribution to the angular momentum loss can be written as the following sum over the initial and final massive states~\cite{DiVecchia:2022owy}:
\begin{equation}\label{eq:staticJ}
	\mathcal{J}^{\alpha\beta} = \frac{G}{2} \sum_{a,b=1}^4 c(\gamma_{ab}) (\eta_a-\eta_b) p_{a}^{[\alpha}p_b^{\beta]}\,,
\end{equation}
where $\eta_a = +1$ ($-1$) if the state is outgoing (incoming), $\eta_{a} p_a$ are the states' momenta,
$\gamma_{ab} = -\eta_a\eta_b p_a\cdot p_b/(m_am_b)$ and 
\begin{equation}
	-c(\gamma) = 
	\frac{\gamma^2-3/2}{\gamma^2-1}
	\,
	\frac{\gamma \operatorname{arccosh}\gamma}{\sqrt{\gamma^2-1}}+\frac{\gamma^2-1/2}{\gamma^2-1}\,.
\end{equation}
Beside this effect related to the \emph{linear} memory, there is another static contribution due to the \emph{nonlinear} memory related to soft gravitons emitted by the gravitational wave itself. Ref.~\cite{Heissenberg:2024umh} has shown that, up to and including $\mathcal{O}(G^4)$, 
nonlinear memory does not contribute\footnote{The contribution due to nonlinear memory was dubbed $\delta \mathcal{J}^{\alpha\beta}$ in~\cite{Heissenberg:2024umh}. See in particular Eq.~(5.72) of that reference for a discussion of why the $\mathcal{O}(G^4)$ term of $b_\alpha \delta \mathcal{J}^{\alpha\beta} p_\beta$ vanishes. A similar mechanism also works for $K_\text{loss}$.} to $\mathcal{J}_\text{stat}$.
The explicit expression for $\mathcal{J}_\text{stat}$ is then
obtained from \eqref{eq:staticJ} by substituting the known results for the impulses $Q_1=p_1+p_4$ and $Q_2=p_2+p_3$ up to $\mathcal{O}(G^3)$ \cite{Herrmann:2021tct}. 

\paragraph*{\textbf{Complete Angular Momentum Loss.}}

Translating the radiative results $\boldsymbol{J}^{\alpha\beta}$ to the center-of-mass frame, $E^{\phantom{\alpha}}_1 \! b_1^\alpha+E^{\phantom{\alpha}}_2 \! b_2^\alpha=0$ with $E_a=-m_a u_a\cdot t$,
then adding the same component of the static loss in this frame
$\mathcal{J}_\text{stat} = b_\alpha \mathcal{J}^{\alpha\beta} p_\beta/(pb)$,
we obtain our final result for $J_\text{loss}$, the total gravitational angular momentum loss as an exact function of $\gamma$ complete to $\mathcal{O}(G^4)$, \texttt{jCM} in the ancillary file.
Similarly \texttt{kCM} is the full 4PM result for the mass dipole component $K_\text{loss}= \tfrac{1}{b} b^\mu J_{\mu\nu} t^\nu$.
With $M=m_1+m_2$ the total mass and $\nu=m_1m_2/M^2$ the symmetric mass ratio,
expanding for small $p_\infty=\sqrt{\gamma^2-1}$ we get
\begin{align}\label{eq:JG4}
\nonumber
&\left.J_{{\rm loss}}\right|_{\mathcal{O}(G^4)}= \frac{G^4 M^5 \nu^2}{b^3 p_{\infty }^2}
\bigg[
\frac{176}{5} 
+
\left(
\frac{8144}{105}-\frac{2984 \nu }{45}
\right)
p_{\infty }^2
\\
&+
\frac{448}{5} p_{\infty }^3
+
\left(
\frac{722 \nu ^2}{9}-\frac{12182 \nu }{225}-\frac{93664}{1575}
\right)
p_\infty^4
\\
&+\left(
\frac{1184}{21}-\frac{220256 \nu }{1575}
\right)
p_\infty^5
+\mathcal{O}(p_\infty^{6})
\bigg],
\nonumber
\end{align}
while the small-velocity expansion of $K_\text{loss}$ reproduces Eq.~(5.80) of \cite{Heissenberg:2024umh}. We also explicitly checked that the expansion of $J_\text{loss}$ agrees with the PN literature~\cite{Bini:2021gat,Bini:2022enm,Heissenberg:2024umh,Geralico:2025rof,Heissenberg:2025fcr,Kavanagh:2026ivj} up to relative $\mathcal{O}(p_\infty^{10})$ for generic $\nu$ and relative $\mathcal{O}(p_\infty^{20})$ in the probe limit.

\paragraph*{\textbf{Conservative 5PM-1SF Scattering Angle.}}

As an important first application of our new 4PM result for $J_{\rm loss}$,
we derive an expression for the 5PM \emph{conservative} scattering angle up to linear order in $\nu$,
i.e.~first-order in self-force (1SF).\footnote{The
	conservative 5PM-1SF angle calculated in \Rcite{Driesse:2024xad}
	has only even-in-velocity contributions.
	We complete this expression with
	the odd-in-velocity terms arising from linear response~\eqref{eq:linresp}.
}
In the approach of \Rcite{Bini:2019nra} the conservative angle is given by:
\begin{equation}
    \label{eq:consPM}
    \theta^\text{cons} = \theta - \theta^\text{rr,lin} + {\mathcal O}(\nu^2)\,,
\end{equation}
where $\theta$ is the full angle,
which at 5PM-1SF was derived in \Rcite{Driesse:2024feo}.
The dissipative term is provided by linear response~\cite{Bini:2012ji}:
\begin{equation}
    \label{eq:linresp}
    \theta^\text{rr,lin}=-\frac12\bigg(\frac{\partial\theta^\text{cons}}{\partial J}J_\text {loss}+
	\frac{\partial\theta^\text{cons}}{\partial E}E_\text{rad} 	\bigg)\,.
\end{equation}
The precise definition of the conservative scattering angle $\theta^\text{cons}$ can be subtle,
but here we are interested in evaluating~\Eqn{eq:linresp} at 5PM order
and so we need to define the conservative dynamics only up to 3PM order,
where there are no ambiguities~\cite{Bern:2019nnu}.
Using the explicit expressions for $E_\text{rad}$ and $J_\text {loss}$ up to 4PM,
we obtain the 5PM expression for \Eqn{eq:linresp} included in the ancillary file. 

We have checked that, expanding the conservative angle~\eqref{eq:consPM} for small velocities,
one reproduces the results of \Rcite{Bini:2025vuk} --- see in particular Eq.~(6.1) of that reference.
From the PM result, it is straightforward to derive also the next subleading terms.
Indicating with $\delta \tilde{\chi}^\text{1cons}_5$ the new corrections to Eq.~(6.1)
starting at 6.5PN we have\footnote{The 5PM coefficient $\tilde{\chi}_5^\text{1cons}$ is defined in \eqref{eq:tildechin} and \eqref{eq:th2ch} following the conventions of~\Rcite{Bini:2025vuk}. The $p_\infty^8$ term was derived in~\cite{Bini:2025zvk} whose result agrees with ours up to a factor of $1/2$ which we believe to be a typo.}
\begin{align}
	\hspace{-30pt}
   \delta&\tilde{\chi}_5^\text{1cons} = \frac{755648}{33075}  p_{\infty}^8  \!+\! \Big[\frac{607416 \log (2 p_\infty)}{13475} - \frac{51696793 \pi ^2}{193536}  
    \nonumber
    \\ &+ \frac{178411037077387}{58677696000}\Big] p_\infty^9
    - \frac{561207424}{5457375} p_{\infty}^{10} + \cdots\,,
\label{eq:ext}
\end{align}
where we focused on the 1SF~${\mathcal O}(\nu)$ term~\eqref{eq:tildechin},
and the dots stand for higher-order PN corrections.

\paragraph*{\textbf{Conclusions.}}

The main result of this Letter is an exact 4PM expression for
the angular momentum lost in a scattering of two Schwarzschild black holes.
This was the last inclusive observable unknown in this setup at 4PM order.
Given that the expression is local in time,
i.e.~it does not contain any $\log p_\infty$ terms signalling tails,
it has immediate relevance for bound orbits --- for example in an
effective-one-body (EOB) resummed description of the dynamics
\cite{Buonanno:1998gg,Khalil:2022ylj,Buonanno:2024vkx,Buonanno:2024byg,Damour:2025uka,Long:2025nmj}.
As we have seen, this missing piece is also a crucial ingredient for
completing the \emph{conservative} description of 5PM dynamics up to first order
in gravitational self-force, $\mathcal{O}(\nu)$.
It will be interesting to see the implications at second order in self-force
once corresponding 5PM scattering results become available.
As for pushing $J_{\rm loss}$ itself to higher PM orders,
the dynamical terms are feasible as long as similar PM calculations
are possible for other dynamical quantities, such as the momentum loss $P_{\rm loss}$.
The static terms will, however, also require further study of contributions
arising from nonlinear memory~\cite{Dlapa:2024cje,Heissenberg:2024umh,Georgoudis:2025vkk,Dlapa:2026oyq}

\paragraph*{\textbf{Note.}}
The \texttt{arXiv} submission of this work was done in coordination with the Humboldt-based group
involving L.~Bohnenblust, C.~J.~Eriksen, G.~U.~Jakobsen,
K.~Stoldt and A.~Yanovets~\cite{berlinJrad},
against whom we have cross-checked our result for $J_{\rm loss}$.

\paragraph*{\textbf{Acknowledgments.}} 
We would like to thank T.~Damour, G.~Uhre Jakobsen and J.~Plefka for valuable discussions.
We are also grateful to the Humboldt group for sharing with us their
complete canonical basis of three-loop master integrals.
This work is supported by The Royal Society grants
URF$\backslash$R$\backslash$221015 and URF$\backslash$R1$\backslash$241371 (A.G.)
and URF\textbackslash R1\textbackslash 231578 (G.M.)
``Gravitational Waves from Worldline Quantum Field Theory''.
R.R.~is supported by the UK EPSRC grant EP/W019663/1, ``CFT and Gravity: Heavy States and Black Holes".
Both G.M.~and R.R.~ are also supported
by the STFC Consolidated Grant ST/X00063X/1, ``Amplitudes, Strings \& Duality''. 
This research utilized Queen Mary's Apocrita HPC facility,
supported by QMUL Research IT~\url{http://doi.org/10.5281/zenodo.438045}.
No new data were generated or analyzed during this study.

\appendix

\section{Notation and Conventions}

We adopt the mostly plus signature for the Minkowski metric $\eta_{\mu\nu}=\text{diag}(-1,+1,\ldots,+1)$.
The initial momenta are regarded as formally outgoing, so they can be expressed in terms of the masses and the initial velocities as 
$
    p_1^\alpha=-m_1^{\phantom{\alpha}} u_1^\alpha
 $, 
 $
    p_2^\alpha=-m_2^{\phantom{\alpha}} u_2^\alpha
$.
The dual  velocities are defined by 
\begin{equation}
        \check u_1^\alpha = \frac{u_1^\alpha-\gamma u_2^\alpha}{\gamma^2-1}\,,
        \quad
        \check u_2^\alpha = \frac{u_2^\alpha-\gamma u_1^\alpha}{\gamma^2-1}\,,
\end{equation}
so that $\check{u}_a\cdot u_b=\delta_{ab}$. 
The spatial momentum of particle 1 in the center-of-mass frame is given by
\begin{equation}\label{eq:pCM}
    p^\alpha = \frac{m_1m_2}{E^2}\left(
    (m_2 + \gamma m_1)u_1^\alpha
    -
    (m_1 + \gamma m_2)u_2^\alpha
    \right).
\end{equation}
We recall the identity
$
    E^2 p^2 = m_1^2 m_2^2 (\gamma^2-1)
$.

The inelastic Fourier transform linking the kernel $W$
to the frequency-domain waveform $\tilde{W}$ is given by
\begin{equation}\label{eq:inelasticFT}
\begin{split}
\tilde{W}^{\mu\nu}&=e^{ib_2\cdot k}\int \frac{d^{D}q_1\,e^{iq_1\cdot b} }{(2\pi)^{D-2}}
\\
&\times
\delta(2p_1\cdot q_1) \delta(2p_2\cdot(q_1+k)) W^{\mu\nu}(q_1,k)\,,
\end{split}
\end{equation} 
where $b^\alpha=b_1^\alpha-b_2^\alpha$.
The elastic Fourier transform linking $\mathbb{P}_\parallel$, $\mathbb{J}_\perp$ to ${P}_\parallel$, ${J}_\perp$ reads
\begin{equation}\label{eq:elasticFT}
\operatorname{FT}{f}=\int \frac{d^{D}q\,e^{iq\cdot b}}{(2\pi)^{D-2}} \delta(2p_1\cdot q) \delta(2p_2\cdot q) f(q)\,.
\end{equation}
For completeness, let us also give the expression for $\mathcal{A}_0$, 
\begin{align}
		\nonumber
			\mathcal{A}_0^{\mu\nu}
			&=8\kappa^{3}m_1^2 m_2^2\Bigg[
			\frac{2\hat{P}_{12}^{\mu} \hat{P}_{12}^{\nu}+\gamma (\hat{P}_{12}^{\mu} \hat{Q}_{12}^{\nu}+\hat{P}_{12}^{\nu} \hat{Q}_{12}^{\mu})}{2q_1^2q_2^2}
			\\
			&+\frac{2\gamma^2-1}8\left(\frac{\hat{Q}_{12}^\mu \hat{Q}_{12}^\nu}{q_1^2q_2^2}-
			\frac{\hat{P}_{12}^\mu \hat{P}_{12}^\nu}{(u_1\cdot k)^2(u_2\cdot k)^2}\right)\Bigg],
            \label{eq:treelevel}
	\end{align}
	where $\kappa=\sqrt{8\pi G}$, while $q_1+q_2+k=0$ and
	\begin{subequations}
		\begin{align}
			\hat{P}_{12}^\mu&=(u_1\cdot k)u_2^\mu- (u_2\cdot k) u_1^\mu\,,\\
			\hat{Q}_{12}^\mu&=(q_1-q_2)^\mu+
			\frac{q_1^2}{u_1\cdot k}u_1^\mu-\frac{q_2^2}{u_2\cdot k}u_2^\mu\,.
		\end{align}
	\end{subequations} 

The conservative deflection angle admits the PM expansion 
\begin{equation}
    \theta^\text{cons} = 
    \frac{E}{M}\sum_{n\ge1} \theta_n^\text{cons} \left(\frac{GM}{b}\right)^n\,,
\end{equation}
where $\theta^\text{cons}_n$ 
are polynomials in $\nu$. At 5PM order
\begin{equation}\label{eq:tildechin}
        \theta_5^\text{cons} = \theta_5^{0\text{cons}} + \nu \theta_5^{1\text{cons}} + \nu^2 \theta_5^{2\text{cons}} \,,
\end{equation}
where $\theta_n^{k\text{cons}}$ are only functions of $\gamma$.
They are closely related to the $\tilde{\chi}_n^\text{1cons}$ and $ \chi_n^\text{cons}$
used in \Rcites{Bini:2025vuk,Bini:2025zvk}:
\begin{equation}
    \label{eq:th2ch}
    \tilde{\chi}_n^\text{1cons} = \tfrac12 \, p_\infty^n \theta_n^\text{1cons}\,,
\end{equation}
and 
\begin{equation}
    \chi_n^\text{cons} = (1-n\nu(\gamma-1))\tilde{\chi}_n^\text{0cons} + \nu \tilde{\chi}_n^\text{1cons}  + \mathcal{O}(\nu^2)\,.
\end{equation}
The full deflection angle admits the same PM expansion:
\begin{equation}
    \theta = 
    \frac{E}{M} \sum_{n\ge1} \theta_n \left(
    \frac{GM}{b}
    \right)^n\,,
\end{equation}
for which \cite{Driesse:2024feo} 
\begin{equation}
    \theta_5 = \theta_5^{0} + \nu \theta_5^{1} + \nu^2 \theta_5^{2} + \left(\frac{M}{E}\right)^2 \nu^3 \theta_5^3\,.
\end{equation}
The 0SF terms are purely conservative, as they are fixed by the geodesic motion, so
\begin{equation}
	\theta_n^{1\text{cons}} = \theta_n^{1} - \theta_n^\text{rr,lin} \,.
\end{equation}
This is because linear response fixes the relation between the conservative and the full deflection angle up to 1SF by
\begin{equation}
    \theta_n = \theta^\text{cons}_n + \theta^\text{rr,lin}_n + \mathcal{O}(\nu^2)\,,
\end{equation}
where, since $\theta^\text{rr,lin}$ starts at $\mathcal{O}(\nu)$,
\begin{equation}
    \theta^\text{rr,lin} = \sum_{n\ge1} {\theta}^\text{rr,lin}_n \left(\frac{GM}{b}\right)^n +\mathcal{O}(\nu^2)\,.
\end{equation}

\begin{widetext}
\section{Functions Appearing in $P_\parallel^\alpha$ and $J_\perp^{\alpha\beta}$}
\label{app:XYZT}

In this section, we present the explicit expressions for the functions in Eq.~\eqref{eq:expl}.
We let ${X}={X}^\text{1rad}+p_\infty {X}^\text{2rad}$ for $X=\mathcal{X}$, $\mathcal{Y}$, $\mathcal{Z}$, $\mathcal{T}$. For the 2rad pieces, we have 
\begin{align}
\begin{aligned}
    \mathcal{X}^\text{2rad}
    &=
   32 \gamma  \left(-12 \gamma ^6+101 \gamma ^2+48\right) \Pi _0^2
   +\frac{16}{15} \left(48 \gamma
   ^8+932 \gamma ^6-1388 \gamma ^4-6947 \gamma ^2-865\right) \Pi _0
   \\
   &\qquad+\frac{16}{45} \gamma  \left(12 \gamma ^6-2444 \gamma ^4+6877 \gamma ^2+7885\right)\,,
\end{aligned}
\end{align}
\begin{align}
\begin{aligned}
    \mathcal{Y}^\text{2rad}
    &=
    +32 \left(4
   \gamma ^8+14 \gamma ^6-56 \gamma ^4-91 \gamma ^2-8\right) \Pi _0^2
   -\frac{64}{15} \gamma  \left(92 \gamma ^6+178 \gamma ^4-1457 \gamma ^2-868\right) \Pi _0\\
   &\qquad+\frac{32}{45} \left(474 \gamma ^6-233 \gamma ^4-5546 \gamma ^2-860\right)\,,
\end{aligned}
\end{align}
\begin{align}
\begin{aligned}
    \mathcal{Z}^\text{2rad}
    &=
   16 \gamma  \left(-12 \gamma ^6+36 \gamma ^4+95 \gamma ^2+18\right) \Pi _0^2
   -\frac{8}{15}
   \left(48 \gamma ^8-764 \gamma ^6+3908 \gamma ^4+4829 \gamma ^2+169\right) \Pi _0
   \\
   &\qquad+\frac{8}{45} \gamma  \left(564 \gamma ^6-2732 \gamma ^4+11347 \gamma ^2+3061\right)\,,
\end{aligned}
\end{align}
\begin{align}
\begin{aligned}
    \mathcal{T}^\text{2rad}
    &=
   16 \left(4 \gamma ^6-100 \gamma ^4-39 \gamma ^2-2\right) \Pi _0^2
   +\frac{8}{5} \gamma  \left(64 \gamma ^6+76 \gamma ^4+2336 \gamma ^2+259\right) \Pi _0
   \\
   &\qquad-\frac{8}{45} \left(972
   \gamma ^6+1456 \gamma ^4+10177 \gamma ^2-320\right)\,,
\end{aligned}
\end{align}
and for the 1rad pieces we have
\begin{align}
    \mathcal{X}^\text{1rad}
    &=
   \frac{1}{16} \gamma  \left(3121 \gamma ^6-1377 \gamma ^4-23537 \gamma ^2-5119\right) \Pi _3
   +\frac{32 \gamma ^2 \left(20 \gamma ^8-36 \gamma ^6-143 \gamma ^4+172 \gamma
   ^2+84\right) \Pi _0^2}{\gamma ^2-1}
   \\
   &-\frac{1}{16} (\gamma -1) \left(2100 \gamma ^7-3946 \gamma ^6-1291 \gamma ^5+5171 \gamma ^4-2511 \gamma ^3+591 \gamma ^2-222 \gamma
   +36\right) \Pi _1
   \nonumber
   \\
   &+\frac{1}{16} (\gamma +1) \left(2100 \gamma ^7+3946 \gamma ^6-1291 \gamma ^5-5171 \gamma ^4-2511 \gamma ^3-591 \gamma ^2-222 \gamma -36\right) \Pi
   _2
   \nonumber
   \\
   &-\frac{\left(344064 \gamma ^{20}+5720752 \gamma ^{18}-14492512 \gamma ^{16}-76682778 \gamma ^{14}+90788740 \gamma ^{12}+36697319 \gamma ^{10}\right) \Pi _0}{6720 \gamma ^9 \left(\gamma ^2-1\right)}
   \nonumber
   \\
   &-\frac{\left(-996450 \gamma ^8+438200 \gamma ^6-123200 \gamma ^4+27300 \gamma ^2-3675\right) \Pi _0}{6720 \gamma ^9 \left(\gamma ^2-1\right)}
   \nonumber
   \\
   &+
   \frac{-3010560 \gamma ^{18}+606428112 \gamma ^{16}-2534929122 \gamma ^{14}-3461843906 \gamma ^{12}+6791018985 \gamma ^{10}}{705600 \gamma ^8 \left(\gamma ^2-1\right)}
   \nonumber
   \\
   &+
   \frac{822250866 \gamma ^8-38739400 \gamma ^6+11230800 \gamma ^4-2609250 \gamma ^2+385875}{705600 \gamma ^8 \left(\gamma ^2-1\right)}
   \nonumber
\end{align} 
\begin{align}
    \mathcal{Y}^\text{1rad}
    &=
    \frac{1}{16} \left(525 \gamma ^8-4711 \gamma ^6-331 \gamma ^4+27351 \gamma ^2+4078\right) \Pi _3
    \\
    &-\frac{32 \gamma  \left(4 \gamma ^{10}+16 \gamma ^8-95 \gamma ^6-49 \gamma ^4+197
   \gamma ^2+24\right) \Pi _0^2}{\gamma ^2-1}
   \nonumber
   \\
   &-\frac{1}{16} (\gamma -1) \left(1050 \gamma ^7-1950 \gamma ^6+916 \gamma ^5+421 \gamma ^4-2801 \gamma ^3+3225 \gamma ^2-1209 \gamma
   +36\right) \Pi _1
   \nonumber
   \\
   &-\frac{1}{16} (\gamma +1) \left(1050 \gamma ^7+1950 \gamma ^6+916 \gamma ^5-421 \gamma ^4-2801 \gamma ^3-3225 \gamma ^2-1209 \gamma -36\right) \Pi
   _2
   \nonumber
   \\
   &+\frac{\left(3076096 \gamma ^{18}+2906996 \gamma ^{16}-63397506 \gamma ^{14}+10564596 \gamma ^{12}+77793073 \gamma ^{10}\right) \Pi _0}{6720 \gamma ^8 \left(\gamma ^2-1\right)}
   \nonumber
   \\
   &+\frac{\left(11209520 \gamma ^8-527100 \gamma ^6+111860 \gamma
   ^4-23450 \gamma ^2+3675\right) \Pi _0}{6720 \gamma ^8 \left(\gamma ^2-1\right)}
   \nonumber
   \\
   &+\frac{-315622912 \gamma ^{16}+270429402 \gamma ^{14}+5643026966 \gamma ^{12}-4271061195 \gamma
   ^{10}}{705600 \gamma ^7 \left(\gamma ^2-1\right)}
   \nonumber
   \\
   &+\frac{-3476793036 \gamma ^8-48652100 \gamma ^6+10309600 \gamma ^4-2205000 \gamma ^2+385875}{705600 \gamma ^7 \left(\gamma ^2-1\right)}
   \nonumber
\end{align}
\begin{align}
    \mathcal{Z}^\text{1rad}
    &=
    \frac{1}{32} \gamma  \left(1321 \gamma ^6-14409 \gamma ^4-24713 \gamma ^2+10889\right) \Pi _3
    +\frac{16 \gamma ^2 \left(20 \gamma ^8-108 \gamma ^6-23 \gamma ^4+214 \gamma
   ^2-6\right) \Pi _0^2}{\gamma ^2-1}
   \\
   &-\frac{1}{64} (\gamma -1) \left(5250 \gamma ^7-14042 \gamma ^6-26597 \gamma ^5+44791 \gamma ^4-7110 \gamma ^3-24042 \gamma ^2+21801 \gamma
   -5811\right) \Pi _1
   \nonumber
   \\
   &+\frac{1}{64} (\gamma +1) \left(5250 \gamma ^7+14042 \gamma ^6-26597 \gamma ^5-44791 \gamma ^4-7110 \gamma ^3+24042 \gamma ^2+21801 \gamma +5811\right) \Pi
   _2
   \nonumber
   \\
   &+\frac{-141496320 \gamma ^{18}+575390532 \gamma ^{16}-2189436492 \gamma ^{14}-2381213096 \gamma ^{12}+5838786795 \gamma ^{10}}{1411200 \gamma ^8 \left(\gamma ^2-1\right)}
   \nonumber
   \\
   &+\frac{513090756 \gamma ^8-44560600 \gamma ^6+10113600
   \gamma ^4-2168250 \gamma ^2+385875}{1411200 \gamma ^8 \left(\gamma ^2-1\right)}
   \nonumber
   \\
   &+\frac{\left(344064 \gamma ^{20}-2747740 \gamma ^{18}+19643686 \gamma ^{16}+50451804 \gamma
   ^{14}-95546122 \gamma ^{12}-15945077 \gamma ^{10}\right) \Pi _0}{13440 \gamma ^9 \left(\gamma
   ^2-1\right)}
   \nonumber
   \\
   &+\frac{\left(2585730 \gamma ^8-486920 \gamma ^6+109760 \gamma ^4-23100 \gamma ^2+3675\right) \Pi _0}{13440 \gamma ^9 \left(\gamma
   ^2-1\right)}
   \nonumber
\end{align}
\begin{align}
    \mathcal{T}^\text{1rad}
    &=
    -\frac{8 \gamma  \left(40 \gamma ^8-338 \gamma ^6+271 \gamma ^4+215 \gamma ^2+6\right) \Pi _0^2}{\gamma ^2-1}
    \\
    &+\frac{1}{32} \left(1050 \gamma ^8-11777 \gamma ^6+12821 \gamma
   ^4+24337 \gamma ^2+481\right) \Pi _3
   \nonumber
   \\
   &-\frac{1}{32} (\gamma -1) \left(2100 \gamma ^7-8100 \gamma ^6-7078 \gamma ^5+11251 \gamma ^4+2589 \gamma ^3-2841 \gamma ^2-759 \gamma
   +318\right) \Pi _1
   \nonumber
   \\
   &-\frac{1}{32} (\gamma +1) \left(2100 \gamma ^7+8100 \gamma ^6-7078 \gamma ^5-11251 \gamma ^4+2589 \gamma ^3+2841 \gamma ^2-759 \gamma -318\right) \Pi
   _2
   \nonumber
   \\
   &+\frac{\left(-1814528 \gamma ^{16}+322632 \gamma ^{14}-68547508 \gamma ^{12}+62527354 \gamma ^{10}\right) \Pi _0}{13440 \gamma ^6 \left(\gamma ^2-1\right)}
   \nonumber
   \\
   &+\frac{\left(47164494 \gamma ^8+1871381 \gamma ^6+166250 \gamma ^4-30100 \gamma
   ^2+4025\right) \Pi _0}{13440 \gamma ^6 \left(\gamma ^2-1\right)}
   \nonumber
   \\
   &+\frac{542236160 \gamma ^{14}-885171436 \gamma ^{12}+6772406658 \gamma ^{10}}{1411200 \gamma ^5 \left(\gamma ^2-1\right)}
   \nonumber
   \\
   &+\frac{-7616856928 \gamma ^8-1010270579
   \gamma ^6+15574650 \gamma ^4-2878750 \gamma ^2+422625}{1411200 \gamma ^5 \left(\gamma ^2-1\right)}
   \nonumber
\end{align}
\end{widetext}

\providecommand{\href}[2]{#2}\begingroup\raggedright\endgroup


\begin{thebibliography}{10}
	
	\bibitem{Bern:2019nnu}
	Z.~Bern, C.~Cheung, R.~Roiban, C.-H. Shen, M.~P. Solon, and M.~Zeng,
	``{Scattering Amplitudes and the Conservative Hamiltonian for Binary Systems
		at Third Post-Minkowskian Order},''
	\href{http://dx.doi.org/10.1103/PhysRevLett.122.201603}{{\em Phys. Rev.
			Lett.} {\bf 122} (2019) no.~20, 201603},
	\href{http://arxiv.org/abs/1901.04424}{{\tt arXiv:1901.04424 [hep-th]}}.
	
	\bibitem{Bern:2019crd}
	Z.~Bern, C.~Cheung, R.~Roiban, C.-H. Shen, M.~P. Solon, and M.~Zeng, ``{Black
		Hole Binary Dynamics from the Double Copy and Effective Theory},''
	\href{http://dx.doi.org/10.1007/JHEP10(2019)206}{{\em JHEP} {\bf 10} (2019)
		206}, \href{http://arxiv.org/abs/1908.01493}{{\tt arXiv:1908.01493
			[hep-th]}}.
	
	\bibitem{Kalin:2020mvi}
	G.~K\"alin and R.~A. Porto, ``{Post-Minkowskian Effective Field Theory for
		Conservative Binary Dynamics},''
	\href{http://dx.doi.org/10.1007/JHEP11(2020)106}{{\em JHEP} {\bf 11} (2020)
		106}, \href{http://arxiv.org/abs/2006.01184}{{\tt arXiv:2006.01184
			[hep-th]}}.
	
	\bibitem{Kalin:2020fhe}
	G.~K\"alin, Z.~Liu, and R.~A. Porto, ``{Conservative Dynamics of Binary Systems
		to Third Post-Minkowskian Order from the Effective Field Theory Approach},''
	\href{http://dx.doi.org/10.1103/PhysRevLett.125.261103}{{\em Phys. Rev.
			Lett.} {\bf 125} (2020) no.~26, 261103},
	\href{http://arxiv.org/abs/2007.04977}{{\tt arXiv:2007.04977 [hep-th]}}.
	
	\bibitem{Mogull:2020sak}
	G.~Mogull, J.~Plefka, and J.~Steinhoff, ``{Classical black hole scattering from
		a worldline quantum field theory},''
	\href{http://dx.doi.org/10.1007/JHEP02(2021)048}{{\em JHEP} {\bf 02} (2021)
		048}, \href{http://arxiv.org/abs/2010.02865}{{\tt arXiv:2010.02865
			[hep-th]}}.
	
	\bibitem{DiVecchia:2020ymx}
	P.~Di~Vecchia, C.~Heissenberg, R.~Russo, and G.~Veneziano, ``{Universality of
		ultra-relativistic gravitational scattering},''
	\href{http://dx.doi.org/10.1016/j.physletb.2020.135924}{{\em Phys. Lett. B}
		{\bf 811} (2020)  135924}, \href{http://arxiv.org/abs/2008.12743}{{\tt
			arXiv:2008.12743 [hep-th]}}.
	
	\bibitem{Damour:2020tta}
	T.~Damour, ``{Radiative contribution to classical gravitational scattering at
		the third order in $G$},''
	\href{http://dx.doi.org/10.1103/PhysRevD.102.124008}{{\em Phys. Rev. D} {\bf
			102} (2020) no.~12, 124008}, \href{http://arxiv.org/abs/2010.01641}{{\tt
			arXiv:2010.01641 [gr-qc]}}.
	
	\bibitem{DiVecchia:2021ndb}
	P.~Di~Vecchia, C.~Heissenberg, R.~Russo, and G.~Veneziano, ``{Radiation
		Reaction from Soft Theorems},''
	\href{http://dx.doi.org/10.1016/j.physletb.2021.136379}{{\em Phys. Lett. B}
		{\bf 818} (2021)  136379}, \href{http://arxiv.org/abs/2101.05772}{{\tt
			arXiv:2101.05772 [hep-th]}}.
	
	\bibitem{Herrmann:2021lqe}
	E.~Herrmann, J.~Parra-Martinez, M.~S. Ruf, and M.~Zeng, ``{Gravitational
		Bremsstrahlung from Reverse Unitarity},''
	\href{http://dx.doi.org/10.1103/PhysRevLett.126.201602}{{\em Phys. Rev.
			Lett.} {\bf 126} (2021) no.~20, 201602},
	\href{http://arxiv.org/abs/2101.07255}{{\tt arXiv:2101.07255 [hep-th]}}.
	
	\bibitem{DiVecchia:2021bdo}
	P.~Di~Vecchia, C.~Heissenberg, R.~Russo, and G.~Veneziano, ``{The eikonal
		approach to gravitational scattering and radiation at $ \mathcal{O}
		(G^{3})$},'' \href{http://dx.doi.org/10.1007/JHEP07(2021)169}{{\em JHEP} {\bf
			07} (2021)  169}, \href{http://arxiv.org/abs/2104.03256}{{\tt
			arXiv:2104.03256 [hep-th]}}.
	
	\bibitem{Herrmann:2021tct}
	E.~Herrmann, J.~Parra-Martinez, M.~S. Ruf, and M.~Zeng, ``{Radiative classical
		gravitational observables at $ \mathcal{O} (G^{3})$ from scattering
		amplitudes},'' \href{http://dx.doi.org/10.1007/JHEP10(2021)148}{{\em JHEP}
		{\bf 10} (2021)  148}, \href{http://arxiv.org/abs/2104.03957}{{\tt
			arXiv:2104.03957 [hep-th]}}.
	
	\bibitem{Bjerrum-Bohr:2021vuf}
	N.~E.~J. Bjerrum-Bohr, P.~H. Damgaard, L.~Plant\'e, and P.~Vanhove,
	``{Classical gravity from loop amplitudes},''
	\href{http://dx.doi.org/10.1103/PhysRevD.104.026009}{{\em Phys. Rev. D} {\bf
			104} (2021) no.~2, 026009}, \href{http://arxiv.org/abs/2104.04510}{{\tt
			arXiv:2104.04510 [hep-th]}}.
	
	\bibitem{Bjerrum-Bohr:2021din}
	N.~E.~J. Bjerrum-Bohr, P.~H. Damgaard, L.~Plant\'e, and P.~Vanhove, ``{The
		amplitude for classical gravitational scattering at third Post-Minkowskian
		order},'' \href{http://dx.doi.org/10.1007/JHEP08(2021)172}{{\em JHEP} {\bf
			08} (2021)  172}, \href{http://arxiv.org/abs/2105.05218}{{\tt
			arXiv:2105.05218 [hep-th]}}.
	
	\bibitem{Brandhuber:2021eyq}
	A.~Brandhuber, G.~Chen, G.~Travaglini, and C.~Wen, ``{Classical gravitational
		scattering from a gauge-invariant double copy},''
	\href{http://dx.doi.org/10.1007/JHEP10(2021)118}{{\em JHEP} {\bf 10} (2021)
		118}, \href{http://arxiv.org/abs/2108.04216}{{\tt arXiv:2108.04216
			[hep-th]}}.
	
	\bibitem{Kalin:2022hph}
	G.~K\"alin, J.~Neef, and R.~A. Porto, ``{Radiation-reaction in the Effective
		Field Theory approach to Post-Minkowskian dynamics},''
	\href{http://dx.doi.org/10.1007/JHEP01(2023)140}{{\em JHEP} {\bf 01} (2023)
		140}, \href{http://arxiv.org/abs/2207.00580}{{\tt arXiv:2207.00580
			[hep-th]}}.
	
	\bibitem{Jakobsen:2022psy}
	G.~U. Jakobsen, G.~Mogull, J.~Plefka, and B.~Sauer, ``{All things retarded:
		radiation-reaction in worldline quantum field theory},''
	\href{http://dx.doi.org/10.1007/JHEP10(2022)128}{{\em JHEP} {\bf 10} (2022)
		128}, \href{http://arxiv.org/abs/2207.00569}{{\tt arXiv:2207.00569
			[hep-th]}}.
	
	\bibitem{Bern:2021dqo}
	Z.~Bern, J.~Parra-Martinez, R.~Roiban, M.~S. Ruf, C.-H. Shen, M.~P. Solon, and
	M.~Zeng, ``{Scattering Amplitudes and Conservative Binary Dynamics at ${\cal
			O}(G^4)$},'' \href{http://dx.doi.org/10.1103/PhysRevLett.126.171601}{{\em
			Phys. Rev. Lett.} {\bf 126} (2021) no.~17, 171601},
	\href{http://arxiv.org/abs/2101.07254}{{\tt arXiv:2101.07254 [hep-th]}}.
	
	\bibitem{Dlapa:2021npj}
	C.~Dlapa, G.~K\"alin, Z.~Liu, and R.~A. Porto, ``{Dynamics of binary systems to
		fourth Post-Minkowskian order from the effective field theory approach},''
	\href{http://dx.doi.org/10.1016/j.physletb.2022.137203}{{\em Phys. Lett. B}
		{\bf 831} (2022)  137203}, \href{http://arxiv.org/abs/2106.08276}{{\tt
			arXiv:2106.08276 [hep-th]}}.
	
	\bibitem{Bern:2021yeh}
	Z.~Bern, J.~Parra-Martinez, R.~Roiban, M.~S. Ruf, C.-H. Shen, M.~P. Solon, and
	M.~Zeng, ``{Scattering Amplitudes, the Tail Effect, and Conservative Binary
		Dynamics at $\mathcal{O}(G^4)$},''
	\href{http://dx.doi.org/10.1103/PhysRevLett.128.161103}{{\em Phys. Rev.
			Lett.} {\bf 128} (2022) no.~16, 161103},
	\href{http://arxiv.org/abs/2112.10750}{{\tt arXiv:2112.10750 [hep-th]}}.
	
	\bibitem{Dlapa:2021vgp}
	C.~Dlapa, G.~K\"alin, Z.~Liu, and R.~A. Porto, ``{Conservative Dynamics of
		Binary Systems at Fourth Post-Minkowskian Order in the Large-Eccentricity
		Expansion},'' \href{http://dx.doi.org/10.1103/PhysRevLett.128.161104}{{\em
			Phys. Rev. Lett.} {\bf 128} (2022) no.~16, 161104},
	\href{http://arxiv.org/abs/2112.11296}{{\tt arXiv:2112.11296 [hep-th]}}.
	
	\bibitem{Dlapa:2022lmu}
	C.~Dlapa, G.~K\"alin, Z.~Liu, J.~Neef, and R.~A. Porto, ``{Radiation Reaction
		and Gravitational Waves at Fourth Post-Minkowskian Order},''
	\href{http://dx.doi.org/10.1103/PhysRevLett.130.101401}{{\em Phys. Rev.
			Lett.} {\bf 130} (2023) no.~10, 101401},
	\href{http://arxiv.org/abs/2210.05541}{{\tt arXiv:2210.05541 [hep-th]}}.
	
	\bibitem{Jakobsen:2023ndj}
	G.~U. Jakobsen, G.~Mogull, J.~Plefka, B.~Sauer, and Y.~Xu, ``{Conservative
		Scattering of Spinning Black Holes at Fourth Post-Minkowskian Order},''
	\href{http://dx.doi.org/10.1103/PhysRevLett.131.151401}{{\em Phys. Rev.
			Lett.} {\bf 131} (2023) no.~15, 151401},
	\href{http://arxiv.org/abs/2306.01714}{{\tt arXiv:2306.01714 [hep-th]}}.
	
	\bibitem{Jakobsen:2023hig}
	G.~U. Jakobsen, G.~Mogull, J.~Plefka, and B.~Sauer, ``{Dissipative Scattering
		of Spinning Black Holes at Fourth Post-Minkowskian Order},''
	\href{http://dx.doi.org/10.1103/PhysRevLett.131.241402}{{\em Phys. Rev.
			Lett.} {\bf 131} (2023) no.~24, 241402},
	\href{http://arxiv.org/abs/2308.11514}{{\tt arXiv:2308.11514 [hep-th]}}.
	
	\bibitem{Damgaard:2023vnx}
	P.~H. Damgaard, E.~R. Hansen, L.~Plant\'e, and P.~Vanhove, ``{The relation
		between KMOC and worldline formalisms for classical gravity},''
	\href{http://dx.doi.org/10.1007/JHEP09(2023)059}{{\em JHEP} {\bf 09} (2023)
		059}, \href{http://arxiv.org/abs/2306.11454}{{\tt arXiv:2306.11454
			[hep-th]}}.
	
	\bibitem{Damgaard:2023ttc}
	P.~H. Damgaard, E.~R. Hansen, L.~Plant\'e, and P.~Vanhove, ``{Classical
		observables from the exponential representation of the gravitational
		S-matrix},'' \href{http://dx.doi.org/10.1007/JHEP09(2023)183}{{\em JHEP} {\bf
			09} (2023)  183}, \href{http://arxiv.org/abs/2307.04746}{{\tt
			arXiv:2307.04746 [hep-th]}}.
	
	\bibitem{Driesse:2024xad}
	M.~Driesse, G.~U. Jakobsen, G.~Mogull, J.~Plefka, B.~Sauer, and J.~Usovitsch,
	``{Conservative Black Hole Scattering at Fifth Post-Minkowskian and First
		Self-Force Order},''
	\href{http://dx.doi.org/10.1103/PhysRevLett.132.241402}{{\em Phys. Rev.
			Lett.} {\bf 132} (2024) no.~24, 241402},
	\href{http://arxiv.org/abs/2403.07781}{{\tt arXiv:2403.07781 [hep-th]}}.
	
	\bibitem{Driesse:2024feo}
	M.~Driesse, G.~U. Jakobsen, A.~Klemm, G.~Mogull, C.~Nega, J.~Plefka, B.~Sauer,
	and J.~Usovitsch, ``{Emergence of Calabi\textendash{}Yau manifolds in
		high-precision black-hole scattering},''
	\href{http://dx.doi.org/10.1038/s41586-025-08984-2}{{\em Nature} {\bf 641}
		(2025) no.~8063, 603--607}, \href{http://arxiv.org/abs/2411.11846}{{\tt
			arXiv:2411.11846 [hep-th]}}.
	
	\bibitem{Driesse:2026qiz}
	M.~Driesse, G.~U. Jakobsen, G.~Mogull, C.~Nega, J.~Plefka, B.~Sauer, and
	J.~Usovitsch, ``{Conservative Black Hole Scattering at Fifth Post-Minkowskian
		and Second Self-Force Order},''
	\href{http://dx.doi.org/10.1103/4dvk-nglx}{{\em Phys. Rev. Lett.} {\bf 137}
		(2026) no.~8, 081402}, \href{http://arxiv.org/abs/2601.16256}{{\tt
			arXiv:2601.16256 [hep-th]}}.
	
	\bibitem{Bern:2024adl}
	Z.~Bern, E.~Herrmann, R.~Roiban, M.~S. Ruf, A.~V. Smirnov, V.~A. Smirnov, and
	M.~Zeng, ``{Amplitudes, supersymmetric black hole scattering at $
		\mathcal{O}({G}^5)$, and loop integration},''
	\href{http://dx.doi.org/10.1007/JHEP10(2024)023}{{\em JHEP} {\bf 10} (2024)
		023}, \href{http://arxiv.org/abs/2406.01554}{{\tt arXiv:2406.01554
			[hep-th]}}.
	
	\bibitem{Bern:2025wyd}
	Z.~Bern, E.~Herrmann, R.~Roiban, M.~S. Ruf, A.~V. Smirnov, S.~Smith, and
	M.~Zeng, ``{Scattering Amplitudes and Conservative Binary Dynamics at
		$\mathcal{O}(G^5)$ without Self-Force Truncation},''
	\href{http://dx.doi.org/10.1103/zjcp-ssrw}{{\em Phys. Rev. Lett.} {\bf 137}
		(2026) no.~2, 021402}, \href{http://arxiv.org/abs/2512.23654}{{\tt
			arXiv:2512.23654 [hep-th]}}.
	
	\bibitem{Veneziano:2022zwh}
	G.~Veneziano and G.~A. Vilkovisky, ``{Angular momentum loss in gravitational
		scattering, radiation reaction, and the Bondi gauge ambiguity},''
	\href{http://dx.doi.org/10.1016/j.physletb.2022.137419}{{\em Phys. Lett. B}
		{\bf 834} (2022)  137419}, \href{http://arxiv.org/abs/2201.11607}{{\tt
			arXiv:2201.11607 [gr-qc]}}.
	
	\bibitem{DiVecchia:2022owy}
	P.~Di~Vecchia, C.~Heissenberg, and R.~Russo, ``{Angular momentum of
		zero-frequency gravitons},''
	\href{http://dx.doi.org/10.1007/JHEP08(2022)172}{{\em JHEP} {\bf 08} (2022)
		172}, \href{http://arxiv.org/abs/2203.11915}{{\tt arXiv:2203.11915
			[hep-th]}}.
	
	\bibitem{Heissenberg:2024umh}
	C.~Heissenberg and R.~Russo, ``{Revisiting gravitational angular momentum and
		mass dipole losses in the eikonal framework},''
	\href{http://dx.doi.org/10.1088/1361-6382/adaabc}{{\em Class. Quant. Grav.}
		{\bf 42} (2025) no.~4, 045014}, \href{http://arxiv.org/abs/2406.03937}{{\tt
			arXiv:2406.03937 [gr-qc]}}.
	
	\bibitem{Georgoudis:2025vkk}
	A.~Georgoudis, V.~Goncalves, C.~Heissenberg, and J.~Parra-Martinez,
	``{Nonlinear Gravitational Memory in the Post-Minkowskian Expansion},''
	\href{http://dx.doi.org/10.1103/8m17-s2y8}{{\em Phys. Rev. Lett.} {\bf 136}
		(2026) no.~12, 121401}, \href{http://arxiv.org/abs/2506.20733}{{\tt
			arXiv:2506.20733 [hep-th]}}.
	
	\bibitem{Manohar:2022dea}
	A.~V. Manohar, A.~K. Ridgway, and C.-H. Shen, ``{Radiated Angular Momentum and
		Dissipative Effects in Classical Scattering},''
	\href{http://dx.doi.org/10.1103/PhysRevLett.129.121601}{{\em Phys. Rev.
			Lett.} {\bf 129} (2022) no.~12, 121601},
	\href{http://arxiv.org/abs/2203.04283}{{\tt arXiv:2203.04283 [hep-th]}}.
	
	\bibitem{DiVecchia:2022piu}
	P.~Di~Vecchia, C.~Heissenberg, R.~Russo, and G.~Veneziano, ``{Classical
		gravitational observables from the eikonal operator},''
	\href{http://dx.doi.org/10.1016/j.physletb.2023.138049}{{\em Phys. Lett. B}
		{\bf 843} (2023)  138049}, \href{http://arxiv.org/abs/2210.12118}{{\tt
			arXiv:2210.12118 [hep-th]}}.
	
	\bibitem{Mogull:2025cfn}
	G.~Mogull, J.~Plefka, and K.~Stoldt, ``{Radiated angular momentum from spinning
		black hole scattering trajectories},''
	\href{http://dx.doi.org/10.1103/my38-14k5}{{\em Phys. Rev. D} {\bf 112}
		(2025) no.~12, 124076}, \href{http://arxiv.org/abs/2506.20643}{{\tt
			arXiv:2506.20643 [hep-th]}}.
	
	\bibitem{Jakobsen:2026ayw}
	G.~U. Jakobsen and K.~Stoldt, ``{A Runway to Dissipation of Angular Momentum
		via Worldline Quantum Field Theory},''
	\href{http://arxiv.org/abs/2605.12214}{{\tt arXiv:2605.12214 [hep-th]}}.
	
	\bibitem{Heissenberg:2025ocy}
	C.~Heissenberg, ``{Radiation-reaction and angular momentum loss at
		$\mathcal{O}(G^4)$},'' \href{http://dx.doi.org/10.1103/xz14-87q7}{{\em Phys.
			Rev. D} {\bf 111} (2025) no.~12, 126012},
	\href{http://arxiv.org/abs/2501.02904}{{\tt arXiv:2501.02904 [hep-th]}}.
	
	\bibitem{Bini:2025zvk}
	D.~Bini, T.~Damour, and A.~Geralico, ``{High-post-Newtonian-order dynamical
		effects induced by tail-of-tail interactions in a two-body system},''
	\href{http://dx.doi.org/10.1103/3jhf-vdjz}{{\em Phys. Rev. D} {\bf 112}
		(2025) no.~4, 044028}, \href{http://arxiv.org/abs/2507.08708}{{\tt
			arXiv:2507.08708 [gr-qc]}}.
	
	\bibitem{Kovacs:1977uw}
	S.~J. Kovacs and K.~S. Thorne, ``{The Generation of Gravitational Waves. 3.
		Derivation of Bremsstrahlung Formulas},''
	\href{http://dx.doi.org/10.1086/155576}{{\em Astrophys. J.} {\bf 217} (1977)
		252--280}.
	
	\bibitem{Kovacs:1978eu}
	S.~J. Kovacs and K.~S. Thorne, ``{The Generation of Gravitational Waves. 4.
		Bremsstrahlung},''
	\href{http://dx.doi.org/10.1086/156350}{{\em Astrophys. J.} {\bf 224} (1978)
		62--85}.
	
	\bibitem{Luna:2017dtq}
	A.~Luna, I.~Nicholson, D.~O'Connell, and C.~D. White, ``{Inelastic Black Hole
		Scattering from Charged Scalar Amplitudes},''
	\href{http://dx.doi.org/10.1007/JHEP03(2018)044}{{\em JHEP} {\bf 03} (2018)
		044}, \href{http://arxiv.org/abs/1711.03901}{{\tt arXiv:1711.03901
			[hep-th]}}.
	
	\bibitem{Jakobsen:2021smu}
	G.~U. Jakobsen, G.~Mogull, J.~Plefka, and J.~Steinhoff, ``{Classical
		Gravitational Bremsstrahlung from a Worldline Quantum Field Theory},''
	\href{http://dx.doi.org/10.1103/PhysRevLett.126.201103}{{\em Phys. Rev.
			Lett.} {\bf 126} (2021) no.~20, 201103},
	\href{http://arxiv.org/abs/2101.12688}{{\tt arXiv:2101.12688 [gr-qc]}}.
	
	\bibitem{Mougiakakos:2021ckm}
	S.~Mougiakakos, M.~M. Riva, and F.~Vernizzi, ``{Gravitational Bremsstrahlung in
		the post-Minkowskian effective field theory},''
	\href{http://dx.doi.org/10.1103/PhysRevD.104.024041}{{\em Phys. Rev. D} {\bf
			104} (2021) no.~2, 024041}, \href{http://arxiv.org/abs/2102.08339}{{\tt
			arXiv:2102.08339 [gr-qc]}}.
	
	\bibitem{DeAngelis:2026kew}
	S.~De~Angelis, V.~F. He, and J.~Parra-Martinez, ``{Analytical Time-Domain
		Waveforms for Black-Hole Scattering in the Post-Minkowskian Expansion},''
	\href{http://arxiv.org/abs/2610.20452}{{\tt arXiv:2610.20452 [gr-qc]}}.
	
	\bibitem{Brandhuber:2023hhy}
	A.~Brandhuber, G.~R. Brown, G.~Chen, S.~De~Angelis, J.~Gowdy, and
	G.~Travaglini, ``{One-loop gravitational bremsstrahlung and waveforms from a
		heavy-mass effective field theory},''
	\href{http://dx.doi.org/10.1007/JHEP06(2023)048}{{\em JHEP} {\bf 06} (2023)
		048}, \href{http://arxiv.org/abs/2303.06111}{{\tt arXiv:2303.06111
			[hep-th]}}.
	
	\bibitem{Herderschee:2023fxh}
	A.~Herderschee, R.~Roiban, and F.~Teng, ``{The sub-leading scattering waveform
		from amplitudes},'' \href{http://dx.doi.org/10.1007/JHEP06(2023)004}{{\em
			JHEP} {\bf 06} (2023)  004}, \href{http://arxiv.org/abs/2303.06112}{{\tt
			arXiv:2303.06112 [hep-th]}}.
	
	\bibitem{Elkhidir:2023dco}
	A.~Elkhidir, D.~O'Connell, M.~Sergola, and I.~A. Vazquez-Holm, ``{Radiation and
		reaction at one loop},''
	\href{http://dx.doi.org/10.1007/JHEP07(2024)272}{{\em JHEP} {\bf 07} (2024)
		272}, \href{http://arxiv.org/abs/2303.06211}{{\tt arXiv:2303.06211
			[hep-th]}}.
	
	\bibitem{Georgoudis:2023lgf}
	A.~Georgoudis, C.~Heissenberg, and I.~Vazquez-Holm, ``{Inelastic exponentiation
		and classical gravitational scattering at one loop},''
	\href{http://dx.doi.org/10.1007/JHEP06(2023)126}{{\em JHEP} {\bf 06} (2023)
		126}, \href{http://arxiv.org/abs/2303.07006}{{\tt arXiv:2303.07006
			[hep-th]}}.
	
	\bibitem{Caron-Huot:2023vxl}
	S.~Caron-Huot, M.~Giroux, H.~S. Hannesdottir, and S.~Mizera, ``{What can be
		measured asymptotically?},''
	\href{http://dx.doi.org/10.1007/JHEP01(2024)139}{{\em JHEP} {\bf 01} (2024)
		139}, \href{http://arxiv.org/abs/2308.02125}{{\tt arXiv:2308.02125
			[hep-th]}}.
	
	\bibitem{Georgoudis:2023ozp}
	A.~Georgoudis, C.~Heissenberg, and I.~Vazquez-Holm, ``{Addendum to: Inelastic
		exponentiation and classical gravitational scattering at one loop},''
	\href{http://dx.doi.org/10.1007/JHEP02(2024)161}{{\em JHEP} {\bf 2024} (2024)
		no.~02, 161}, \href{http://arxiv.org/abs/2312.14710}{{\tt arXiv:2312.14710
			[hep-th]}}.
	
	\bibitem{Georgoudis:2023eke}
	A.~Georgoudis, C.~Heissenberg, and R.~Russo, ``{An eikonal-inspired approach to
		the gravitational scattering waveform},''
	\href{http://dx.doi.org/10.1007/JHEP03(2024)089}{{\em JHEP} {\bf 03} (2024)
		089}, \href{http://arxiv.org/abs/2312.07452}{{\tt arXiv:2312.07452
			[hep-th]}}.
	
	\bibitem{Georgoudis:2024pdz}
	A.~Georgoudis, C.~Heissenberg, and R.~Russo, ``{Post-Newtonian multipoles from
		the next-to-leading post-Minkowskian gravitational waveform},''
	\href{http://dx.doi.org/10.1103/PhysRevD.109.106020}{{\em Phys. Rev. D} {\bf
			109} (2024) no.~10, 106020}, \href{http://arxiv.org/abs/2402.06361}{{\tt
			arXiv:2402.06361 [hep-th]}}.
	
	\bibitem{Bini:2024rsy}
	D.~Bini, T.~Damour, S.~De~Angelis, A.~Geralico, A.~Herderschee, R.~Roiban, and
	F.~Teng, ``{Gravitational waveforms: A tale of two formalisms},''
	\href{http://dx.doi.org/10.1103/PhysRevD.109.125008}{{\em Phys. Rev. D} {\bf
			109} (2024) no.~12, 125008}, \href{http://arxiv.org/abs/2402.06604}{{\tt
			arXiv:2402.06604 [hep-th]}}.
	
	\bibitem{Brunello:2025eso}
	G.~Brunello, S.~De~Angelis, and D.~A. Kosower, ``{Analytic one-loop scattering
		waveform in general relativity},''
	\href{http://dx.doi.org/10.1103/6xtl-qn8z}{{\em Phys. Rev. D} {\bf 114}
		(2026) no.~4, 045004}, \href{http://arxiv.org/abs/2511.05412}{{\tt
			arXiv:2511.05412 [hep-th]}}.
	
	\bibitem{Bini:2026dvn}
	D.~Bini, T.~Damour, and A.~Geralico, ``{Quadrupolar bremsstrahlung waveform at
		the third-and-a-half post-Newtonian accuracy},''
	\href{http://arxiv.org/abs/2604.21522}{{\tt arXiv:2604.21522 [gr-qc]}}.
	
	\bibitem{Bini:2026jwc}
	D.~Bini, T.~Damour, and A.~Geralico, ``{Octupolar bremsstrahlung waveform up to
		the two-loop level and the third-and-a-half post-Newtonian accuracy},''
	\href{http://arxiv.org/abs/2609.11835}{{\tt arXiv:2609.11835 [gr-qc]}}.
	
	\bibitem{Anastasiou:2002qz}
	C.~Anastasiou, L.~J. Dixon, and K.~Melnikov, ``{NLO Higgs boson rapidity
		distributions at hadron colliders},''
	\href{http://dx.doi.org/10.1016/S0920-5632(03)80168-8}{{\em Nucl. Phys. B
			Proc. Suppl.} {\bf 116} (2003)  193--197},
	\href{http://arxiv.org/abs/hep-ph/0211141}{{\tt arXiv:hep-ph/0211141}}.
	
	\bibitem{Anastasiou:2003yy}
	C.~Anastasiou, L.~J. Dixon, K.~Melnikov, and F.~Petriello, ``{Dilepton rapidity
		distribution in the Drell-Yan process at NNLO in QCD},''
	\href{http://dx.doi.org/10.1103/PhysRevLett.91.182002}{{\em Phys. Rev. Lett.}
		{\bf 91} (2003)  182002}, \href{http://arxiv.org/abs/hep-ph/0306192}{{\tt
			arXiv:hep-ph/0306192}}.
	
	\bibitem{Anastasiou:2015yha}
	C.~Anastasiou, C.~Duhr, F.~Dulat, E.~Furlan, F.~Herzog, and B.~Mistlberger,
	``{Soft expansion of double-real-virtual corrections to Higgs production at
		N$^{3}$LO},'' \href{http://dx.doi.org/10.1007/JHEP08(2015)051}{{\em JHEP}
		{\bf 08} (2015)  051}, \href{http://arxiv.org/abs/1505.04110}{{\tt
			arXiv:1505.04110 [hep-ph]}}.
	
	\bibitem{He:2025how}
	V.~F. He and J.~Parra-Martinez, ``{Generalized Unitarity Method for Worldline
		Field Theory},'' \href{http://arxiv.org/abs/2510.00989}{{\tt arXiv:2510.00989
			[hep-th]}}.
	
	\bibitem{Henn:2013pwa}
	J.~M. Henn, ``{Multiloop integrals in dimensional regularization made
		simple},'' \href{http://dx.doi.org/10.1103/PhysRevLett.110.251601}{{\em Phys.
			Rev. Lett.} {\bf 110} (2013)  251601},
	\href{http://arxiv.org/abs/1304.1806}{{\tt arXiv:1304.1806 [hep-th]}}.
	
	\bibitem{Lee:2012cn}
	R.~N. Lee, ``{Presenting LiteRed: a tool for the Loop InTEgrals REDuction},''
	\href{http://arxiv.org/abs/1212.2685}{{\tt arXiv:1212.2685 [hep-ph]}}.
	
	\bibitem{Lee:2013mka}
	R.~N. Lee, ``{LiteRed 1.4: a powerful tool for reduction of multiloop
		integrals},'' \href{http://dx.doi.org/10.1088/1742-6596/523/1/012059}{{\em J.
			Phys. Conf. Ser.} {\bf 523} (2014)  012059},
	\href{http://arxiv.org/abs/1310.1145}{{\tt arXiv:1310.1145 [hep-ph]}}.
	
	\bibitem{Maierhofer:2017gsa}
	P.~Maierh{\"o}fer, J.~Usovitsch, and P.~Uwer, ``{Kira{\textemdash}A Feynman
		integral reduction program},''
	\href{http://dx.doi.org/10.1016/j.cpc.2018.04.012}{{\em Comput. Phys.
			Commun.} {\bf 230} (2018)  99--112},
	\href{http://arxiv.org/abs/1705.05610}{{\tt arXiv:1705.05610 [hep-ph]}}.
	
	\bibitem{Klappert:2020nbg}
	J.~Klappert, F.~Lange, P.~Maierh{\"o}fer, and J.~Usovitsch, ``{Integral
		reduction with Kira 2.0 and finite field methods},''
	\href{http://dx.doi.org/10.1016/j.cpc.2021.108024}{{\em Comput. Phys.
			Commun.} {\bf 266} (2021)  108024},
	\href{http://arxiv.org/abs/2008.06494}{{\tt arXiv:2008.06494 [hep-ph]}}.
	
	\bibitem{Lange:2025fba}
	F.~Lange, J.~Usovitsch, and Z.~Wu, ``{Kira 3: integral reduction with efficient
		seeding and optimized equation selection},''
	\href{http://dx.doi.org/10.1016/j.cpc.2025.109999}{{\em Comput. Phys.
			Commun.} {\bf 322} (2026)  109999},
	\href{http://arxiv.org/abs/2505.20197}{{\tt arXiv:2505.20197 [hep-ph]}}.
	
	\bibitem{Heissenberg:2025fcr}
	C.~Heissenberg and R.~Russo, ``{Reconstructing the gravitational waveform from
		its probe limit},''
	\href{http://dx.doi.org/10.1016/j.nuclphysb.2026.117538}{{\em Nucl. Phys. B}
		{\bf 1029} (2026)  117538}, \href{http://arxiv.org/abs/2511.13835}{{\tt
			arXiv:2511.13835 [hep-th]}}.
	
	\bibitem{Riva:2023xxm}
	M.~M. Riva, F.~Vernizzi, and L.~K. Wong, ``{Angular momentum balance in
		gravitational two-body scattering: Flux, memory, and supertranslation
		invariance},'' \href{http://dx.doi.org/10.1103/PhysRevD.108.104052}{{\em
			Phys. Rev. D} {\bf 108} (2023) no.~10, 104052},
	\href{http://arxiv.org/abs/2302.09065}{{\tt arXiv:2302.09065 [gr-qc]}}.
	
	\bibitem{Bini:2021gat}
	D.~Bini, T.~Damour, and A.~Geralico, ``{Radiative contributions to
		gravitational scattering},''
	\href{http://dx.doi.org/10.1103/PhysRevD.104.084031}{{\em Phys. Rev. D} {\bf
			104} (2021) no.~8, 084031}, \href{http://arxiv.org/abs/2107.08896}{{\tt
			arXiv:2107.08896 [gr-qc]}}.
	
	\bibitem{Bini:2022enm}
	D.~Bini, T.~Damour, and A.~Geralico, ``{Radiated momentum and radiation
		reaction in gravitational two-body scattering including time-asymmetric
		effects},'' \href{http://dx.doi.org/10.1103/PhysRevD.107.024012}{{\em Phys.
			Rev. D} {\bf 107} (2023) no.~2, 024012},
	\href{http://arxiv.org/abs/2210.07165}{{\tt arXiv:2210.07165 [gr-qc]}}.
	
	\bibitem{Geralico:2025rof}
	A.~Geralico, ``{Scattering of a point mass by a Schwarzschild black hole:
		Radiated energy and angular momentum},''
	\href{http://dx.doi.org/10.1103/ntd5-h858}{{\em Phys. Rev. D} {\bf 112}
		(2025) no.~6, 064088}, \href{http://arxiv.org/abs/2507.03442}{{\tt
			arXiv:2507.03442 [gr-qc]}}.
	
	\bibitem{Kavanagh:2026ivj}
	C.~Kavanagh, J.~Parra-Martinez, and D.~Usseglio, ``{Analytical Time-Domain
		Waveforms for Black-Hole Scattering in the Self-Force Expansion},''
	\href{http://arxiv.org/abs/2610.20451}{{\tt arXiv:2610.20451 [gr-qc]}}.
	
	\bibitem{Bini:2019nra}
	D.~Bini, T.~Damour, and A.~Geralico, ``{Novel approach to binary dynamics:
		application to the fifth post-Newtonian level},''
	\href{http://dx.doi.org/10.1103/PhysRevLett.123.231104}{{\em Phys. Rev.
			Lett.} {\bf 123} (2019) no.~23, 231104},
	\href{http://arxiv.org/abs/1909.02375}{{\tt arXiv:1909.02375 [gr-qc]}}.
	
	\bibitem{Bini:2012ji}
	D.~Bini and T.~Damour, ``{Gravitational radiation reaction along general orbits
		in the effective one-body formalism},''
	\href{http://dx.doi.org/10.1103/PhysRevD.86.124012}{{\em Phys. Rev. D} {\bf
			86} (2012)  124012}, \href{http://arxiv.org/abs/1210.2834}{{\tt
			arXiv:1210.2834 [gr-qc]}}.
	
	\bibitem{Bini:2025vuk}
	D.~Bini and T.~Damour, ``{High precision black hole scattering: Tutti frutti vs
		worldline effective field theory},''
	\href{http://dx.doi.org/10.1103/8ks7-2blq}{{\em Phys. Rev. D} {\bf 112}
		(2025) no.~4, 044002}, \href{http://arxiv.org/abs/2504.20204}{{\tt
			arXiv:2504.20204 [hep-th]}}.
	
	\bibitem{Buonanno:1998gg}
	A.~Buonanno and T.~Damour, ``{Effective one-body approach to general
		relativistic two-body dynamics},''
	\href{http://dx.doi.org/10.1103/PhysRevD.59.084006}{{\em Phys. Rev.} {\bf
			D59} (1999)  084006},
	\href{http://arxiv.org/abs/gr-qc/9811091}{{\tt arXiv:gr-qc/9811091 [gr-qc]}}.
	
	\bibitem{Khalil:2022ylj}
	M.~Khalil, A.~Buonanno, J.~Steinhoff, and J.~Vines, ``{Energetics and
		scattering of gravitational two-body systems at fourth post-Minkowskian
		order},'' \href{http://dx.doi.org/10.1103/PhysRevD.106.024042}{{\em Phys.
			Rev. D} {\bf 106} (2022) no.~2, 024042},
	\href{http://arxiv.org/abs/2204.05047}{{\tt arXiv:2204.05047 [gr-qc]}}.
	
	\bibitem{Buonanno:2024vkx}
	A.~Buonanno, G.~U. Jakobsen, and G.~Mogull, ``{Post-Minkowskian theory meets
		the spinning effective-one-body approach for two-body scattering},''
	\href{http://dx.doi.org/10.1103/PhysRevD.110.044038}{{\em Phys. Rev. D} {\bf
			110} (2024) no.~4, 044038}, \href{http://arxiv.org/abs/2402.12342}{{\tt
			arXiv:2402.12342 [gr-qc]}}.
	
	\bibitem{Buonanno:2024byg}
	A.~Buonanno, G.~Mogull, R.~Patil, and L.~Pompili, ``{Post-Minkowskian Theory
		Meets the Spinning Effective-One-Body Approach for Bound-Orbit Waveforms},''
	\href{http://dx.doi.org/10.1103/PhysRevLett.133.211402}{{\em Phys. Rev.
			Lett.} {\bf 133} (2024) no.~21, 211402},
	\href{http://arxiv.org/abs/2405.19181}{{\tt arXiv:2405.19181 [gr-qc]}}.
	
	\bibitem{Damour:2025uka}
	T.~Damour, A.~Nagar, A.~Placidi, and P.~Rettegno, ``{Novel Lagrange-multiplier
		approach to the effective-one-body dynamics of binary systems in
		post-Minkowskian gravity},'' \href{http://dx.doi.org/10.1103/41sd-g2gb}{{\em
			Phys. Rev. D} {\bf 113} (2026) no.~2, 024042},
	\href{http://arxiv.org/abs/2503.05487}{{\tt arXiv:2503.05487 [gr-qc]}}.
	
	\bibitem{Long:2025nmj}
	O.~Long, H.~P. Pfeiffer, A.~Buonanno, G.~U. Jakobsen, G.~Mogull,
	A.~Ramos-Buades, H.~R. R{\"u}ter, L.~E. Kidder, and M.~A. Scheel, ``{Highly
		accurate simulations of asymmetric black-hole scattering and cross validation
		of effective-one-body models},''
	\href{http://dx.doi.org/10.1103/p6fx-7798}{{\em Phys. Rev. D} {\bf 112}
		(2025) no.~12, 124039}, \href{http://arxiv.org/abs/2507.08071}{{\tt
			arXiv:2507.08071 [gr-qc]}}.
	
	\bibitem{Dlapa:2024cje}
	C.~Dlapa, G.~K{\"a}lin, Z.~Liu, and R.~A. Porto, ``{Local in Time Conservative
		Binary Dynamics at Fourth Post-Minkowskian Order},''
	\href{http://dx.doi.org/10.1103/PhysRevLett.132.221401}{{\em Phys. Rev.
			Lett.} {\bf 132} (2024) no.~22, 221401},
	\href{http://arxiv.org/abs/2403.04853}{{\tt arXiv:2403.04853 [hep-th]}}.
	
	\bibitem{Dlapa:2026oyq}
	C.~Dlapa, G.~K{\"a}lin, Z.~Liu, and R.~A. Porto, ``{Nonlocal-in-time tail
		effects in gravitational scattering to fifth post-Minkowskian and tenth
		self-force orders},'' \href{http://dx.doi.org/10.1103/wkp4-vy6g}{{\em Phys.
			Rev. D} {\bf 114} (2026) no.~2, 024029},
	\href{http://arxiv.org/abs/2604.25916}{{\tt arXiv:2604.25916 [hep-th]}}.
	
	\bibitem{berlinJrad}
	L.~Bohnenblust, C.~J. Eriksen, G.~U. Jakobsen, K.~Stoldt, and A.~Yanovets,
	``Mechanical angular momentum loss at $\mathcal{O}({G}^4)$ from static and
	radiative gravitons.'' To appear.
	
\end{thebibliography}
\end{document}